\documentclass{emulateapj}
\usepackage{amsmath}
\usepackage{graphicx}
\usepackage{natbib}

\received{July 31, 2019}
\revised{August}
\accepted{August}

\shorttitle{Triangulum--Andromeda overdensity}
\shortauthors{Sales Silva et al.}

\begin{document}

\title{Triangulum-Andromeda overdensity: a region with a complex stellar population}


\author{J. V. Sales Silva\altaffilmark{1}, H. D. Perottoni\altaffilmark{2,3}, K. Cunha\altaffilmark{1,5}, H. J. Rocha-Pinto\altaffilmark{2}, F. Almeida-Fernandes\altaffilmark{2,3},Diogo Souto\altaffilmark{1},S. R. Majewski\altaffilmark{6}}
\altaffiltext{1}{Observat\'orio Nacional/MCTIC, R. Gen. Jos\'e Cristino, 77,  20921-400, Rio de Janeiro, Brazil}
\altaffiltext{2}{Universidade Federal do Rio de Janeiro, Observat\'orio do Valongo, Lad. Pedro Ant\^onio 43, 20080-090, Rio de Janeiro, Brazil}
\altaffiltext{3}{Departamento de Astronomia, IAG, Universidade de S\~ao Paulo, Rua do Mat\~ao, 1226, 05509-900, S\~ao Paulo, Brazil}
\altaffiltext{4}{Steward Observatory, University of Arizona Tucson AZ 85719} 
\altaffiltext{5}{Departamento de F\'isica, Universidade Federal de Sergipe, Av. Marechal Rondon, S/N, 49000-000 S\~ao Crist\'ov\~ao, SE, Brazil}
\altaffiltext{6}{Department of Astronomy, University of Virginia, Charlottesville, VA 22904-4325, USA}

\begin{abstract}

The Triangulum--Andromeda (TriAnd) overdensity is a distant structure of the Milky Way located in the second Galactic quadrant well below the Galactic plane. 
Since its discovery, its nature has been under discussion, whether it could be old perturbations of the Galactic disk or the remains of a disrupted former dwarf galaxy. 
In this study, we investigate the kinematics and chemical composition in 13 stars selected as TriAnd candidates from 2MASS photometry. The sample was observed using the GRACES high-resolution spectrograph attached to the Gemini North telescope. Based on radial velocities obtained from the spectra and the astrometric data from GAIA, three different kinematic criteria were used to classify our sample stars as belonging to the TriAnd overdensity. The TriAnd confirmed members in our sample span a range in metallicities, including two metal-poor stars ([Fe/H] $\sim$ -1.3 dex).
We show that the adopted kinematical classification also chemically segregates TriAnd and non-TriAnd members of our sample, 
indicating a unique chemical pattern of the TriAnd stars. 
Our results indicate different chemical patterns for the [Na/Fe], [Al/Fe], [Ba/Fe], and [Eu/Fe] ratios in the TriAnd stars when compared to the chemical pattern of the local disk; the paucity of studies chemically characterizing the outer disk population of the Milky Way is the main obstacle in establishing that the TriAnd population is chemically similar to field stars in the outer disk. But TriAnd chemical pattern is reminiscent of that found in outer disk open clusters, although the latter are significantly more metal-rich than TriAnd.

\end{abstract}


\section{Introduction} 
\label{sec:intro}

In the last decades, the level of detail and detection of structure in the Galaxy has intensified considerably with the advance of the observational power of the telescopes and their instruments, and the completion of large surveys, which enables the discoveries of dwarf galaxies (\citealt{Willman2005a,Belokurov2007b, DrlicaWagner2015, Koposov2015, Laevens2015,kimjerjen2015}), stellar streams (e.g., \citealt{Newberg2002}; \citealt{Majewski2003SagStream}; \citealt{Grillmair2006}; \citealt{bernard2016}; \citealt{malhan2018}; \citealt{shipp2018}; \citealt{Perottoni2019}), and stellar overdensities (e.g., \citealt{RP2004}, \citealt{Majewski2004}; \citealt{Martin2007}; \citealt{Juric2008}; \citealt{belokurov_hercules_2007}; \citealt{Sesar2007}, \citealt{Watkins2009}; \citealt{Li2016}). 
One of these overdensities is Triangulum--Andromeda (TriAnd), which was discovered by \citet{RP2004} and \citet{Majewski2004}. \citet{RP2004} found TriAnd from a sample of 2MASS M giant candidates while searching for asymmetries in the sky-projected stellar density, and \citet{Majewski2004} located the main-sequence and the main-sequence turnoff of the TriAnd overdensity using data from a deep photometric survey developed to study the stellar halo of M 31.

The TriAnd overdensity is very tenuous and extends to approximately 1000 square degrees covering $100^{\circ} < l < 150^{\circ}$ and $-15^{\circ} > b > -35^{\circ}$ (\citealt{RP2004,Deason2014,Sheffield2014,Perottoni2018}), in the southern Galactic hemisphere. Estimates for the heliocentric distance of the TriAnd population range between 15--21 kpc \citep{Sheffield2014, Martin2014}, its age spans a range between 6--10 Gyr \citep{Sheffield2014} and it has stars that go $\sim$7 kpc below the galactic plane (\citealt{Hayes2018}). Moreover, the TriAnd stellar population seems to move in orbits nearly coplanar to the disk with a low velocity dispersion (\citealt{RP2004,Sheffield2014}) and it is possibly associated with Mon/GASS and A13 in the velocity space (\citealt{Li2017, Sheffield2018}).

As other overdensities seen at low $b$, TriAnd is subject to many hypotheses that suggest it to be part of the outer Galactic disk (\citealt{Xu2015,Price2015,Li2017,Sheffield2018,Hayes2018,Bergemann2018}) or of a disrupted dwarf galaxy (\citealt{Chou2011,Sheffield2014,Deason2014}). 
The ratio of RR Lyrae stars to M giant stars in the TriAnd region is compatible with that of the Galactic disk and not with the typical ratio for a dwarf galaxy \citep{Price2015}. But this could be a consequence of its metallicity, for TriAnd stellar population is more metal-rich than that of known dwarf galaxies --- present TriAnd [Fe/H] estimates range from $-0.4$ to $-1.3$ dex (\citealt{Deason2014, Sheffield2014, Hayes2018, Bergemann2018, FernandezAlvar2019}). 

\citet{Chou2011} performed the first high-resolution study of six stars belonging to TriAnd. They obtained that three of their targets had [Ti/Fe] in agreement with the disk pattern, while for three others their [Ti /Fe] was lower than the disk value; they concluded that TriAnd could have an extragalactic origin.
More recently, \citet{Bergemann2018} and \citet{Hayes2018} using optical and infrared high-resolution spectroscopy, respectively, found support for a Galactic origin for the TriAnd stars, despite their large height below the Galactic plane ($\approx7$ kpc, \citealt{Hayes2018}). Both studies attributed their odd location to tidal interactions of the disk with passing or merging dwarf galaxies. This hypothesis is corroborated by several theoretical works on N-Body and hydrodynamical simulations (\citealt{Purcell2011, Gomez2013,Gomez2016,Laporte2018a, Laporte2018b}).

\citet{Bergemann2018} presented  chemical abundances of O, Na, Mg, Ti, Fe, Ba, and Eu, finding that TriAnd stars have an abundance pattern similar to that 
of Galactic disk stars, while \citet{Hayes2018}, using chemical abundances results from the APOGEE survey from DR14 (\citealt{Majewski2017}) for Fe, C, N, Mg, K, Ca, Mn and Ni concluded that TriAnd stars are similar to outer disk stars, but with a metallicity ([Fe/H] $\approx -0.8$ dex) lower than most of those with R$_{GC} <$ 15 kpc. 

Considering the interest in further understanding the nature of the TriAnd overdensity, in this study we conduct a kinematic and chemical study of a sample of thirteen TriAnd candidate stars observed with high-resolution spectroscopy to investigate their chemo-kinematical properties. We determined the chemical abundances of seven elements (Na, Al, Fe, Cr, Ni, Ba and Eu) for all the stars in our sample; moreover, using astrometric data from Gaia, we derived the kinematic properties and orbits of the stars.

\begin{figure}
\includegraphics[width=\columnwidth]{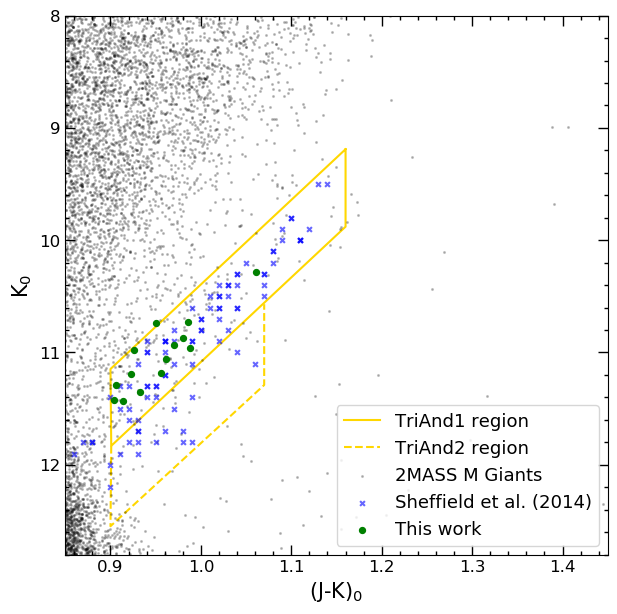}
\caption{Color--magnitude diagram of 2MASS M giants candidates located between 90$^{\circ} < l < 160^{\circ}$ and $-10^{\circ} > b > -40^{\circ}$. The green circles show the TriAnd candidates analysed in this work. The blue crosses show the TriAnd1 and TriAnd2 M giants candidates sample from \citet{Sheffield2014}. The yellow parallelograms having the solid and dashed borders are, respectively, approximate selection regions for TriAnd1 and TriAnd2 members by \citet{Sheffield2014}.}%
\label{fig:CMD_2MASS}
\end{figure}

\section{Sample selection and observations} \label{sec:sample}

A TriAnd candidate star sample was initially defined by \citet{RP2004} from the 2MASS catalog \citep{Cutri2003} through a color criterion in the $J-H$ vs. $J-K$ diagram that segregates M giants from M dwarfs. It was the same criterion employed by \citet{Majewski2003SagStream} and \citet{RP2003} for mapping the Sagittarius dSph tidal tails and the Monoceros stream, respectively. \citet{RP2004} used a statistical method to estimate the most likely distance of each star, such that the TriAnd candidate sample was formed by those M giants having $100^{\degr} < l < 150^{\degr}$, $-50^{\degr} < b < -15^{\degr}$ and most likely distances between 15 and 30 kpc from the Sun\footnote{Recently, \citet{Sheffield2014} showed that \citeauthor{RP2004}'s selection included stars from both TriAnd and TriAnd2, which was discovered a few years later by \citet{Martin2007}.}. 

We overimposed this M giant TriAnd sample to the TriAnd map made by \citet{Perottoni2018}, which is based on SDSS main-sequence stars, to randomly select our spectroscopic program stars. These were picked around the coordinates of three of the densest regions in the Perottoni et al.'s map since our goal was to study whether there were chemical differences between those three regions. The final sample selected for the spectroscopic program according to color--magnitude selection areas from \citet{Sheffield2014} is likely to be formed by a mix of TriAnd stars (see Figure \ref{fig:CMD_2MASS}) and a few interloping halo giants that may lay in the line of sight. We deredden the data the 2MASS following the Equations (1) from \citet{Majewski2003SagStream}.

The observations of thirteen TriAnd candidate stars were carried out in the second half of 2016 (two exposures for each star) using the GRACES (Gemini Remote Access to CFHT ESPaDOnS Spectrograph) instrument (\citealt{Tollestrup2012}) attached to the Gemini North telescope located in Hawaii/USA. 
The data were reduced using the OPERA pipeline (\citealt{Martioli2012}) and included: bias subtraction, flat-field correction, and wavelength calibration; we used the IRAF package to perform spectra normalization. Our spectra have high-resolution (R = 40,000) and a S/N $\sim$ 50-70 per \AA at 6000 \AA. The wavelength range of the spectra is from $\sim$ 4000 to 10,000 \AA. 
We determined the radial velocities of our sample using the task {\tt fxcor} in IRAF to cross-correlate the observed spectra with templates from \citet{Munari2005}. In Table \ref{table:observed_candidates} we show relevant information about the observed stars: star 2MASS identification, equatorial and galactic coordinates, $J$, $H$, $K$, $J-K$ magnitudes from 2MASS. 
In addition, we obtained from the ESO Archive the spectrum of the star 2M23174139+3113043 of the sample of \citet{Bergemann2018}. This star was observed in the Very Large Telescope using the UVES spectrograph. For detailed information about this spectrum see \citet{Bergemann2018}. We added this star to our sample for a comparison between the results in the two studies.

\begin{table*} 
\tabcolsep 0.20truecm
\caption{TriAnd candidates observed with GRACES.}
\begin{tabular}{llccccccccc}\hline\hline
 & 2MASS ID           &      RA       &     DEC        &    $l$   &    $b$   &    $J$   &    $H$   &    $K$   &  $K_{0}$   &   $(J-K)_{0}$  \\\hline
\# &                   &    hh:mm:ss   &    dd:mm:ss    & degree & degree &  mag   &  mag   &  mag   &   mag   \\\hline 
1 & 00075751+3359414   &  00:07:57.510 &  33:59:41.400  & 112.731& $-$28.014& 12.352 & 11.556 & 11.420 & 11.405 & 0.904 \\
2 & 00534976+4626089   &  00:53:49.760 &  46:26:08.900  & 123.362& $-$16.434& 11.914 & 11.032 & 10.867 & 10.833 & 0.980 \\
3 & 00594094+4614332   &  00:59:40.940 &  46:14:33.200  & 124.419& $-$16.605& 12.203 & 11.381 & 11.183 & 11.149 & 0.956\\
4 & 01020943+4643251   &  01:02:09.430 &  46:43:25.100  & 124.844& $-$16.108& 12.268 & 11.480 & 11.292 & 11.256 & 0.906\\
5 & 01151944+4713512   &  01:15:19.440 &  47:13:51.200  & 127.135& $-$15.446& 12.165 & 11.337 & 11.190 & 11.163 & 0.923\\
6 & 02485891+4312154   &  02:48:58.910 &  43:12:15.400  & 144.630& $-$14.660& 11.398 & 10.506 & 10.284 & 10.256 & 1.061 \\
7 & 23535441+3449575   &  23:53:54.410 &  34:49:57.500  & 109.755& $-$26.563& 11.944 & 11.152 & 10.975 & 10.953 & 0.926 \\
8 & 23481637+3129372   &  23:48:16.370 &  31:29:37.200  & 107.473& $-$29.475& 12.080 & 11.274 & 11.059  & 11.028 & 0.961\\
9 & 02350813+4455263   &  02:35:08.130 &  44:55:26.300  & 141.550& $-$14.166& 12.403 & 11.674 & 11.435  & 11.407 & 0.914\\
10 & 23495808+3445569   &  23:49:58.080 &  34:45:56.900  & 108.863& $-$26.421& 11.744 & 10.896 & 10.729 & 10.714 & 0.86 \\
11 & 02510349+4342045   &  02:51:03.490 &  43:42:04.500  & 144.743& $-$14.046& 11.999 & 11.117 & 10.957 & 10.929 & 0.987\\
12 & 02475442+4429269   &  02:47:54.420 &  44:29:26.900  & 143.856& $-$13.597& 11.972 & 11.105 & 10.931 &10.895 & 0.971\\
13 & 02463235+4314481   &  02:46:32.350 &  43:14:48.200  & 144.199& $-$14.821& 12.335 & 11.556 & 11.351 & 11.324 & 0.932\\
14 & 23174139+3113043$^{a}$   &  23:17:41.390 &  31:13:04.300  & 100.379& $-$27.515& 11.750 & 10.920 & 10.740  & 10.701 & 0.950 \\\hline
\hline
\end{tabular}
\\Notes. Columns, from left to right: star 2MASS identification, R.A., decl., galactic coordinates ($l$ and $b$),\\ $J$, $H$, $K$, $J-K$ from 2MASS, $K_{0}$, $(J-K)_{0}$.\\ $^{a}$ Spectrum from ESO-ARCHIVE.
\label{table:observed_candidates}
\end{table*}

\section{Atmospheric parameters and abundance analysis} \label{sec:atm}

The first step was to define the line list and the model atmospheres that will be used to determine the chemical abundances and the atmospheric parameters of the stars in our sample. 

\begin{figure*}[htbp]
    \begin{minipage}{0.47\linewidth}
        \centering
        \includegraphics[width=3.0in]{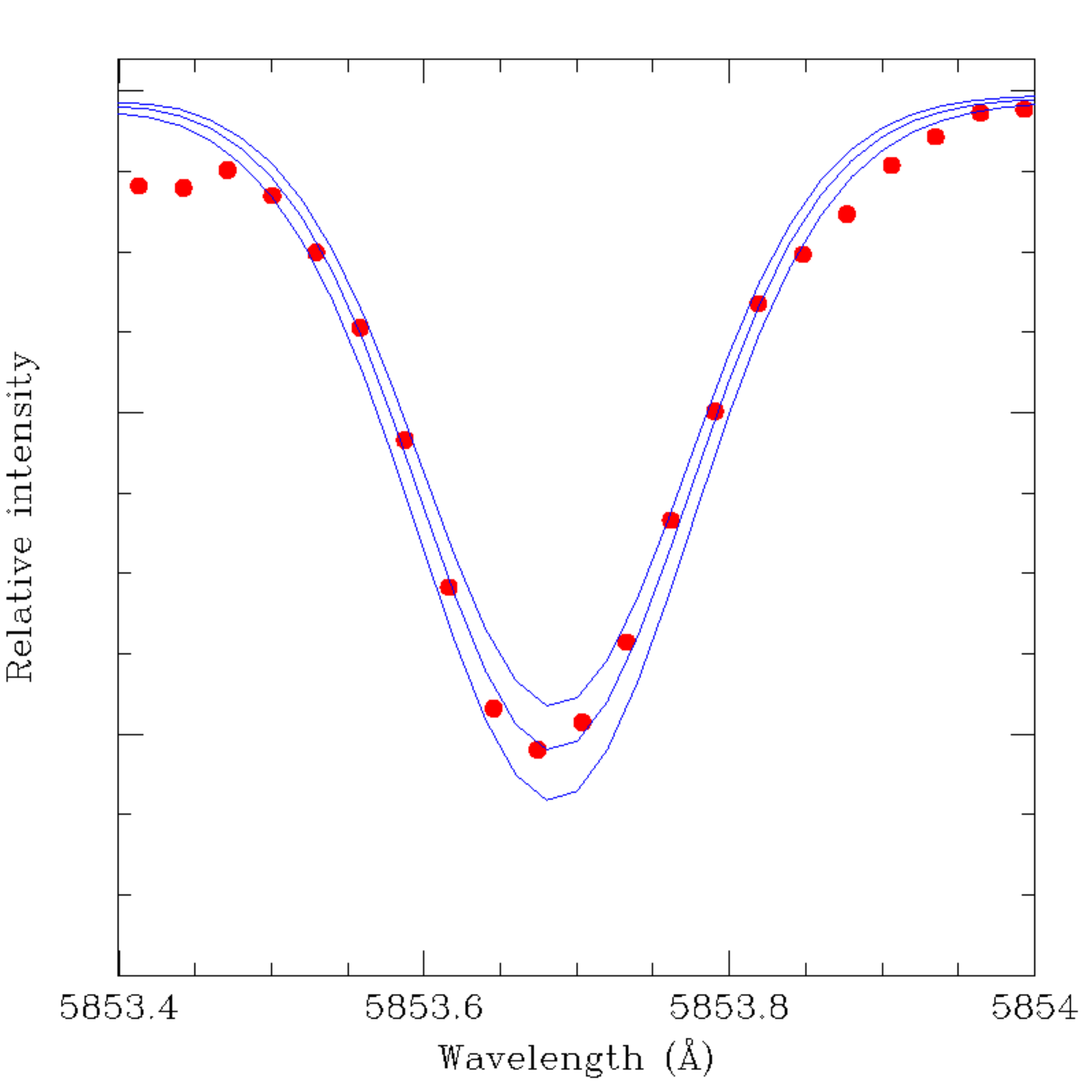}\\
    \end{minipage}
    \begin{minipage}{0.47\linewidth}
        \centering
        \includegraphics[width=3.0in]{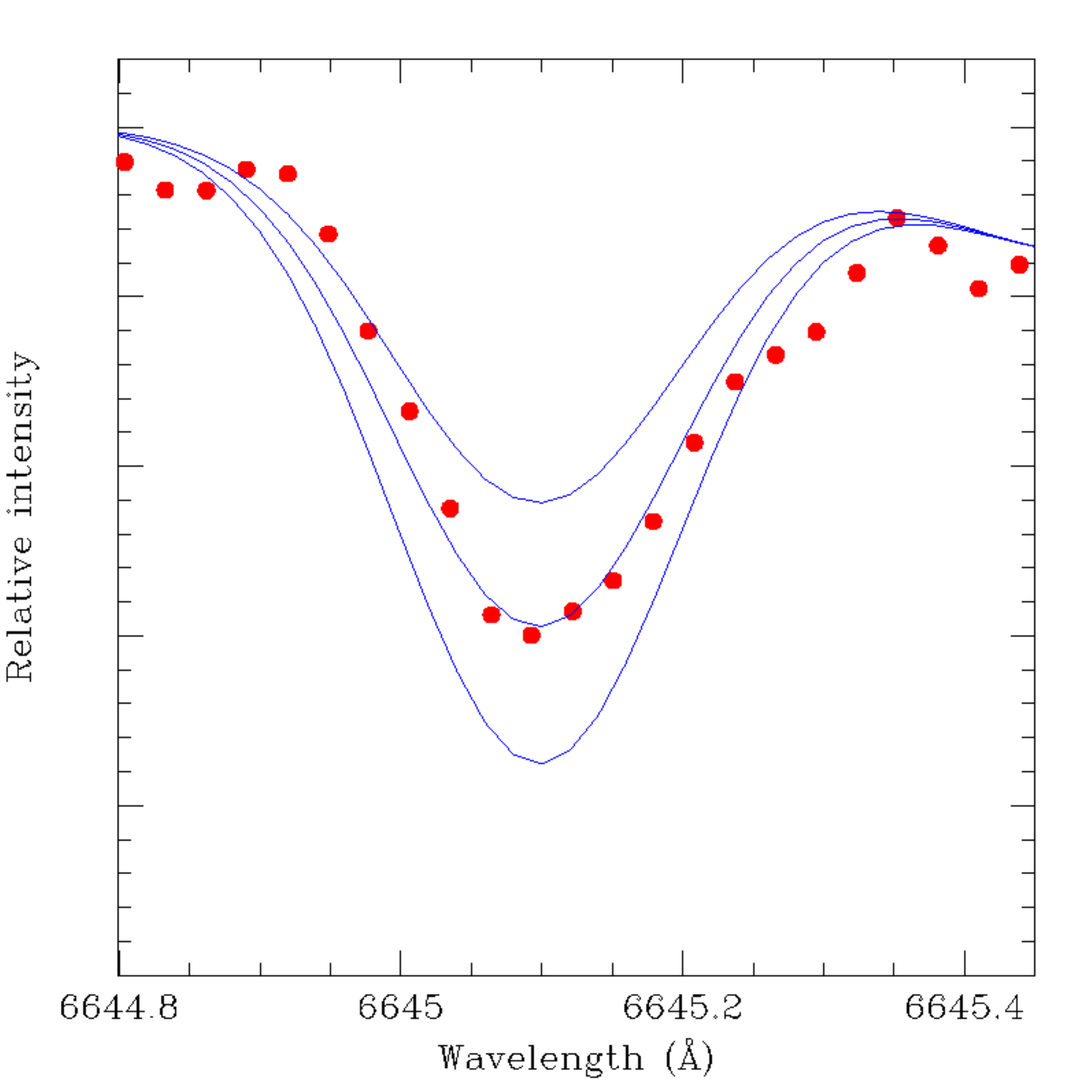}\\
    \end{minipage}
    \caption{Observed (dotted red line) and synthetic spectra (solid blue lines) in the region of the Ba II and Eu II lines at 5853 \AA{} and 6645 \AA{}, respectively, for the star 7. The synthetic spectra in the 5853 \AA{} region represents [Ba/Fe]= 0.26, 0.46 and 0.66 whereas for the 6645 \AA{} region the synthetic spectra represents [Eu/Fe]= 0.20, 0.40 and 0.60.}
    \label{sinteBaEu}
\end{figure*}

Our list of Fe I and Fe II absorption lines is composed of about 170 lines, with the atomic parameters obtained from \citet{Heiter2015}.
To analyze the neutral and ionized Fe lines, we carefully looked whether the lines overlapped with other absorption features; thus only the unblended Fe absorption lines were used in our analysis. 
We measured the equivalent widths of the lines of the elements Fe, Na, Al, Cr and Ni using the task {\tt splot} in IRAF. We note, however, that due to overlapping absorption lines it was not possible to determine chromium abundances for the entire sample. In the Appendix, we show the measurements of equivalent widths used to obtain chemical abundances.
For the neutron capture elements (Ba and Eu) we used the spectral synthesis method due to the hyperfine structure and the contribution of different isotopes to the absorption lines. To determine the abundances of Ba and Eu we used the absorption lines of Ba II at 5853 \AA\ and Eu II at 6645 \AA, respectively. The Ba II line at 5853 \AA\ was selected due to the small NLTE corrections obtained for this transition (\citealt{Andrievsky2013}; \citealt{Bergemann2018}). 
In Figure \ref{sinteBaEu} we show for the target star 7 an example of the spectral syntheses and best fit abundances for the Ba II line (left panel) and the Eu II line (right panel). 
The atomic parameters of the iron-peak elements, as well as Na, Al and Ba were obtained from \citet{Heiter2015}, \citet{SalesSilva2016} and \citet{McWilliam1998}, whereas for Eu we used the atomic parameters from M. Roriz et al. (2019, in preparation). For Cr and Na we adopted the NLTE corrections from \citet{Bergemann2014} and \citet{Lind2011}, respectively. 
For Al I transitions analyzed, the NLTE corrections are negligible
(\citealt{Nordlander2017}). 
We note that for Ni there were no NLTE studies available in the literature.

We adopted the spherical MARCS models (\citealt{Gustafsson2008}) and used the MOOG code (\citealt{Sneden1973}) to perform 1-D LTE analysis and generate synthetic spectra as well as compute abundances from equivalent widths (EW).  For determination of the atmospheric parameters, we used the excitation, and ionization equilibrium approaches for the Fe I and Fe II absorption lines, besides the independence of the equivalent width with the abundance of Fe I. We also did test calculations with Kurucz plane-parallel models (\citealt{Kurucz1996}). In general, the stellar parameter results using the different atmospheric models (MARCS and Kurucz) were very similar with a mean difference within the uncertainties of the respective parameters: $<\Delta$T$_{\rm eff(MARCS-kurucz)}>$ = 49 K, $<\Delta$logg$_{\rm MARCS-kurucz}>$ = 0.1, $\Delta\xi_{\rm MARCS-kurucz}$ = 0.08 km/s and $<\Delta$[Fe/H]$_{\rm MARCS-kurucz}>$ = 0.04. We selected the MARCS models to define the final atmospheric parameters of our stars knowing that this choice would not influence our results and conclusions because both models present very similar results. In Table \ref{table:parameters} we show the atmospheric parameters adopted for the stars of our sample. 
In the Appendix, we show the uncertainties associated with atmospheric parameters and chemical abundances.

As previously mentioned, for the sake of comparison we added one star observed by \citet{Bergemann2018} to our sample. The atmospheric parameters obtained for this star (\#14) are also in Table \ref{table:parameters}. Our results are very similar to those obtained by \citet{Bergemann2018} for T$_{\rm eff}$, $\xi$, and [Fe/H], only for log $g$ we found a large discrepancy ($\Delta$ log $g$ = 0.72), with our log $g$ being smaller than that determined in \citet{Bergemann2018}. As in this study, \citet{Bergemann2018} determined log $g$ through the ionization equilibrium of Fe~I and Fe~II. 
To further test our methodology, we determined the atmospheric parameters for the Sun and the 'standard' red giant Arcturus applying the same methodology used here for the TriAnd candidate stars, and we found values very similar to those found in the literature (see Appendix), including log $g$.

\begin{table*} 
\tabcolsep 0.24truecm
\caption{Derived atmospheric parameters, radial velocities, and metallicity for TriAnd stars candidates. For [Fe\,{\sc i}/H] and 
[Fe\,{\sc ii}/H] we also show the standard deviation and the number of lines employed.}
\begin{tabular}{lccccccccc}\hline\hline
&  $T_{\rm eff}$     & log $g$        & $\xi$                & [Fe\,{\sc i}/H]$\pm$ $\sigma$ (\#) & [Fe\,{\sc ii}/H]$\pm$ $\sigma$ (\#) &    RV         & $\mu_\alpha \ddagger$ & $\mu_\delta \ddagger$  &     Membership $\dagger$    \\\hline
\# &     K              &                & km\,s$^{-1}$         &                                    &                                     & km\,s$^{-1}$     & mas/yr   & mas/yr &                  \\\hline 
1 &  4150              &  0.6           &  0.88                &  $-$1.71$\pm$0.09\,(49)            & $-$1.70$\pm$0.08\,(7)               &  $-$233.6$\pm$0.5&  0.83$\pm$0.04	&	-0.45$\pm$0.02  &  Non-Member     \\
2 &  3925              &  1.3           &  1.81                &  $-$0.46$\pm$0.10\,(39)            & $-$0.46$\pm$0.08\,(5)               &  $-$245.3$\pm$1.5&  -0.74$\pm$0.07 & 0.12$\pm$0.05  &Non-Member     \\
3 &  4100              &  0.4           &  1.77                &  $-$0.82$\pm$0.10\,(48)            & $-$0.81$\pm$0.10\,(5)               &  $-$165.6$\pm$0.5&  -0.15$\pm$0.06 & 0.06$\pm$0.05  &Member        \\
4 &  4125              &  0.0           &  1.04                &  $-$1.50$\pm$0.11\,(58)            & $-$1.49$\pm$0.12\,(6)               &  $-$144.0$\pm$0.8&  1.88$\pm$0.05 & -0.76$\pm$0.05  &Non-Member     \\
5 &  4075              &  0.7           &  1.99                &  $-$0.94$\pm$0.10\,(49)            & $-$0.95$\pm$0.16\,(5)               &  $-$162.4$\pm$0.5&  -0.29$\pm$0.08 & -0.02$\pm$0.09  &Member        \\
6 &  3900              &  0.5           &  1.82                &  $-$0.81$\pm$0.13\,(38)            & $-$0.80$\pm$0.07\,(5)               &  $-$139.5$\pm$1.4&  0.20$\pm$0.05 & -0.20$\pm$0.05  &Member        \\
7 &  4200              &  1.3           &  0.60                &  $-$0.78$\pm$0.10\,(49)            & $-$0.78$\pm$0.04\,(5)               &  $-$118.6$\pm$0.6&  -0.44$\pm$0.03 & -0.72$\pm$0.03  &Member        \\
8 &  3975              &  0.4           &  1.93                &  $-$1.42$\pm$0.08\,(52)            & $-$1.40$\pm$0.08\,(5)               &  $-$196.9$\pm$0.6&  1.20$\pm$0.04 & -0.99$\pm$0.03  &Non-Member     \\
9 &  4050              &  0.5           &  0.59                &  $-$1.23$\pm$0.09\,(44)            & $-$1.22$\pm$0.10\,(5)               &  $-$139.7$\pm$0.9&  0.18$\pm$0.06 & -0.17$\pm$0.06  &Member        \\
10 &  3925              &  1.4           &  1.96                &  $-$0.63$\pm$0.10\,(42)            & $-$0.62$\pm$0.16\,(4)               &   $-$43.4$\pm$0.6&  1.75$\pm$0.04 & -0.49$\pm$0.02  &Non-Member     \\
11 &  4025              &  0.9           &  1.63                &  $-$0.78$\pm$0.10\,(53)            & $-$0.77$\pm$0.10\,(8)               &  $-$106.2$\pm$0.7&  -0.06$\pm$0.07 & -0.23$\pm$0.05  &  Member        \\
12 &  4000              &  0.3           &  0.97                &  $-$1.34$\pm$0.12\,(45)            & $-$1.34$\pm$0.13\,(3)               &  $-$145.2$\pm$1.0&  0.17$\pm$0.05 & -0.26$\pm$0.05  & Member        \\
13 &  4100              &  0.6           &  1.98                &  $-$1.03$\pm$0.08\,(46)            & $-$1.03$\pm$0.15\,(7)               &  $-$182.1$\pm$0.5&  1.45$\pm$0.07 & -1.22$\pm$0.06  &Non-Member     \\
14 &  3925              &  0.3           &  1.62                &  $-$0.91$\pm$0.13\,(54)            & $-$0.89$\pm$0.12\,(6)               &  $-$85.1$\pm$0.5&  1.29$\pm$0.05 &	-0.97$\pm$0.04  & Non-Member     \\\hline
\hline
\end{tabular}
\\$\dagger$ See Subsection \ref{sec:kinematics}.
\\$\ddagger$ Proper motion obtained from Gaia DR2 catalog (See Subsection \ref{sec:kinematics}).
\label{table:parameters}
\end{table*}

\section{Results} \label{sec:resu}

\subsection{Kinematics of TriAnd stars}
\label{sec:kinematics}

\begin{figure}
\includegraphics[width=\columnwidth]{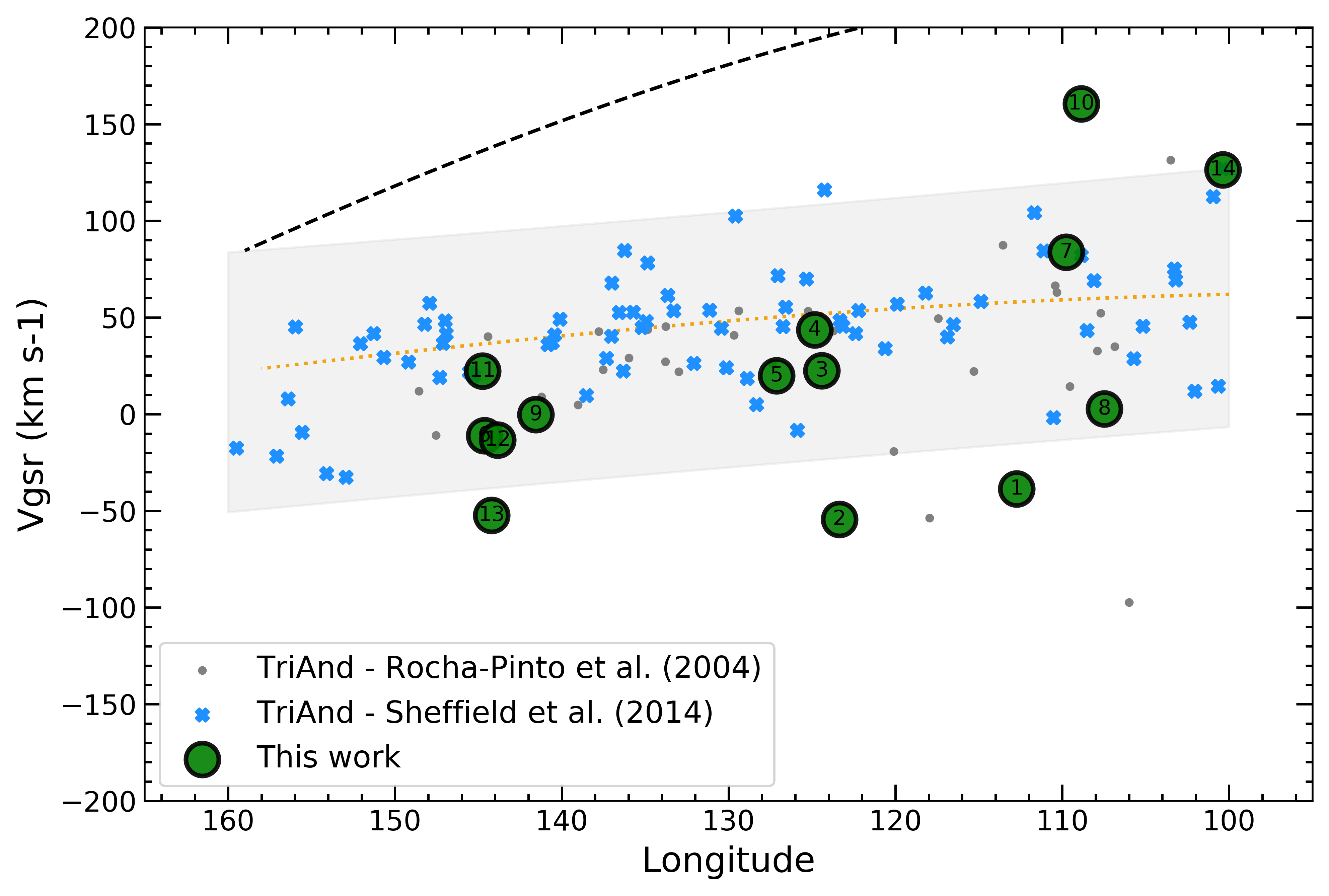}
\caption{Radial velocity in the Galactocentric Standard of Rest as a function of Galactic longitude for the TriAnd candidates. The symbols in green, blue, and grey show respectively the TriAnd candidates from this work, \citet{Sheffield2014}, and \citet{RP2004}. The dashed black line and the orange dotted curve correspond to the circular velocity\textsuperscript{$\ddagger$} for local stars and stars at R$_{GC}$ = 30 kpc, respectively. The grey shaded solid region show 2$\sigma$ prediction interval where the TriAnd sample would be confined. The numbers inside the green circles follow in order top-to-bottom the stars in Table \ref{table:parameters}.  The stars 1, 2, 10, 13 are outside the grey area, region that indicates where the TriAnd population lie. \\ \textsuperscript{$\ddagger$} The rotation speed adopted is $\Theta_{0}$ = 236 km s$^{-1}$ \citep{Bovy2009} and the peculiar motion of the Sun relative to the LSR as given in \citet{Schonrich2010}.}
\label{fig:vgsr_long}
\end{figure}

The sample selection criteria used in the previous section were based on spatial position and a color--magnitude cut. However, to better identify TriAnd candidates we also apply kinematic criteria. As in \citet{Sheffield2014}, we used the radial velocity in the Galactocentric Standard of Rest as a function of Galactic longitude as sample criterion (see Figure \ref{fig:vgsr_long}). 
We used the samples from \citet{RP2004} and \citet{Sheffield2014} to obtain the 2$\sigma$ prediction interval where a star from TriAnd overdensity should lie. According to this criterion the stars \#1, \#2, \#10, and \#13 are not members of TriAnd.

\begin{figure}
\includegraphics[width=\columnwidth]{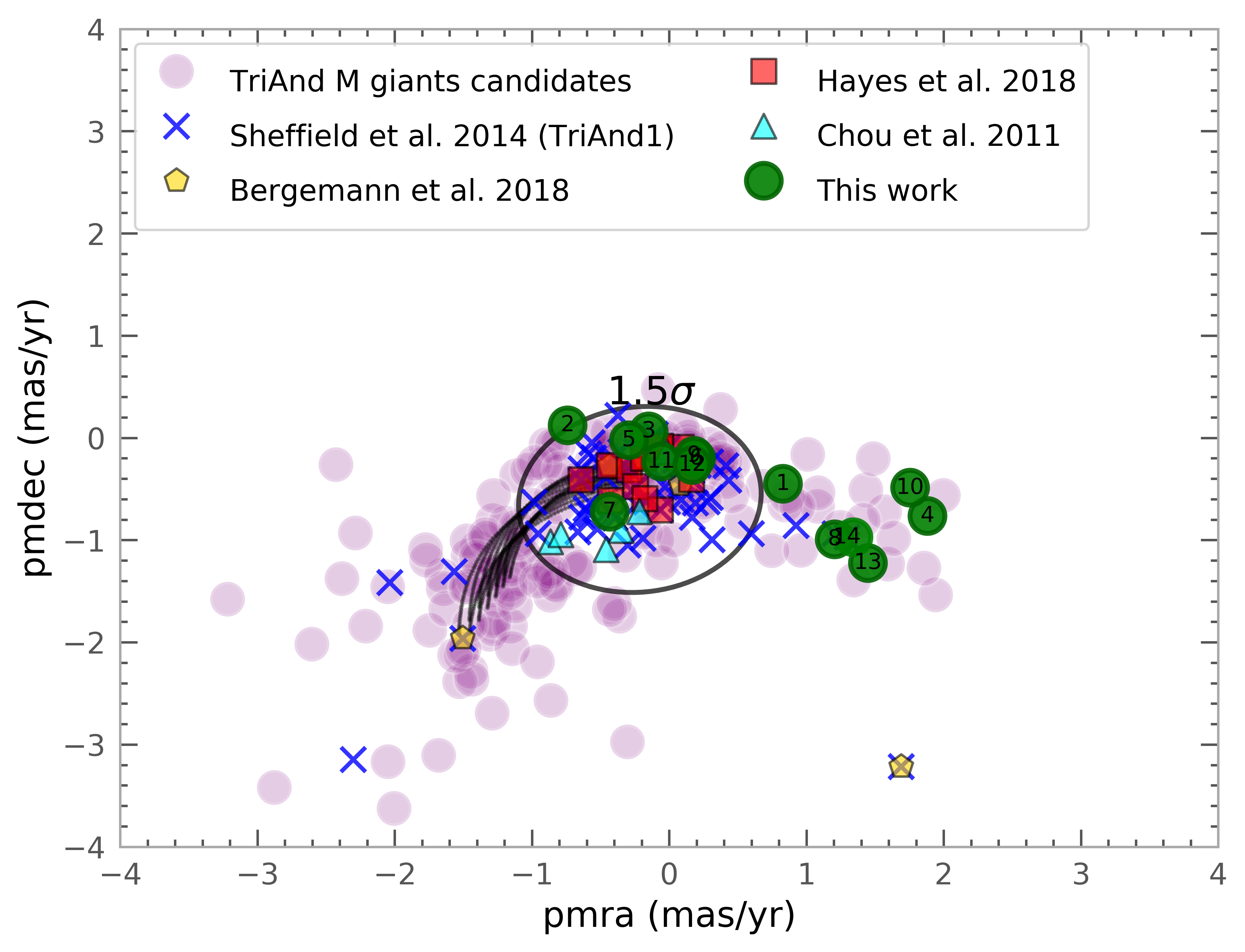}
\caption{GAIA proper motion of TriAnd candidates stars. The blue crosses, green circles, yellow pentagon, red squares, and the cyan triangles, are respectively the TriAnd samples of \citet{Sheffield2014}, this work, \citet{Bergemann2018}, \citet{Hayes2018}, and \citet{Chou2011}. The purple circles are TriAnd M giants candidates without spectroscopic observation selected in the color-magnitude boxes from \citet{Sheffield2014} as describe in Section \ref{sec:sample}. The numbers inside the green circles follow in order top-to-bottom the stars in Table \ref{table:parameters}. The black dots indicate the expected proper motion for two stellar populations with heliocentric distances between 18--23 kpc with Z = -5 and -7 kpc. The 1.5$\sigma$ ellipsoid indicates the characteristic proper motion of those samples. The stars 1, 2, 4, 8, 10, 13 are outside of the ellipsoid which indicates the characteristic proper motion of TriAnd candidates}%
\label{fig:proper_motion}
\end{figure}

We cross-matched the sample from this work, \citet{Sheffield2014}, \citet{Bergemann2018}, \citet{Hayes2018}, and \citet{Chou2011} with the Gaia DR2 catalog \citep{gaia_mission, gaia_dr2} to obtain the proper motion of TriAnd candidate stars. Figure \ref{fig:proper_motion} shows the proper motion of TriAnd candidates stars. We used a 1.5$\sigma$ ellipsoid around the centroid of the proper motion distribution of TriAnd candidates to estimate the TriAnd characteristic proper motion. The stars \#1, \#2, \#4, \#8, \#10, and \#13 are outside of the ellipsoid and were classified as non-member according to their proper motion. 
We call attention to the fact that the stars located in the region of the green circles \#1, \#4, \#8, \#10, \#13 in Figure \ref{fig:proper_motion} have a randomly spatial distribution in the TriAnd region (below we discuss the orbit of these stars) while the proper motion of the other stars are varying with the longitude.

We used the Astropy library to obtain the expected proper motion for two stellar population with circular orbits situated at heliocentric distances between 18--23 kpc with $Z = -$5 and $-$7 kpc. These objects are distributed between 90$^{\circ} < l < 160^{\circ}$ and $-15^{\circ} > b \approx -30^{\circ}$. Figure \ref{fig:proper_motion} shows that the expected proper motion of objects located at the position of TriAnd is compatible with the proper motion of TriAnd candidates. 

As a further test to our selection criteria, we have also classified the stars according to their orbital parameters. In order to do this, we have used the stellar positions, GAIA DR2 proper motions, our measured radial velocities and our estimated distances, to integrate the stellar orbits applying the \texttt{Galpy} integrator \citep{Bovy2015}. Distances were roughly estimated by photometric parallax. The orbits of all 14 stars in our sample can be seen in Figure \ref{fig:orbits} in Appendix \ref{appendix:orbits}. We have also integrated the orbits of the \citet{Bergemann2018} and \citet{Hayes2018} samples using the radial velocities and distances provided in both studies.

We characterize the orbits by defining the {\em eccentricity} and {\em orbital diskness} from the estimated perigalactic and apogalactic radius ($R_\mathrm{peri}$, $R_\mathrm{apo}$) and maximum distance from the galactic plane ($z_{\mathrm{max}}$):

\begin{equation}
    {\rm eccentricity} = \frac{R_\mathrm{apo} - R_\mathrm{peri}}{R_\mathrm{apo} + R_\mathrm{peri}}
\label{eq:ecc}
\end{equation}
\begin{equation}
    {\rm orbital\,diskness} = \frac{R_\mathrm{apo} - z_\mathrm{max}}{R_\mathrm{apo} + z_\mathrm{max}}
\label{eq:od}
\end{equation}

By definition, the eccentricity ranges from 0 to 1 and measures how circular the projected orbit is in the galactic plane: 0 corresponds to a perfectly circular orbit, while 1 corresponds to a straight line. Similarly, the ``orbital diskness'' ranges from -1 to 1 and measures how confined is the stellar orbit to the galactic disk: where 1 corresponds to an orbit perfectly confined to the plane of the disk, 0 to an orbit where the vertical motion has a range equivalent to the range of the motion in the disk, and -1 when the vertical motion is much greater than the motion in the disk (a case which is very unlikely to happen to stellar orbits). These definitions allow us to characterize the stellar orbits and classify groups who display similar parameters.

\begin{figure}
\includegraphics[width=\columnwidth]{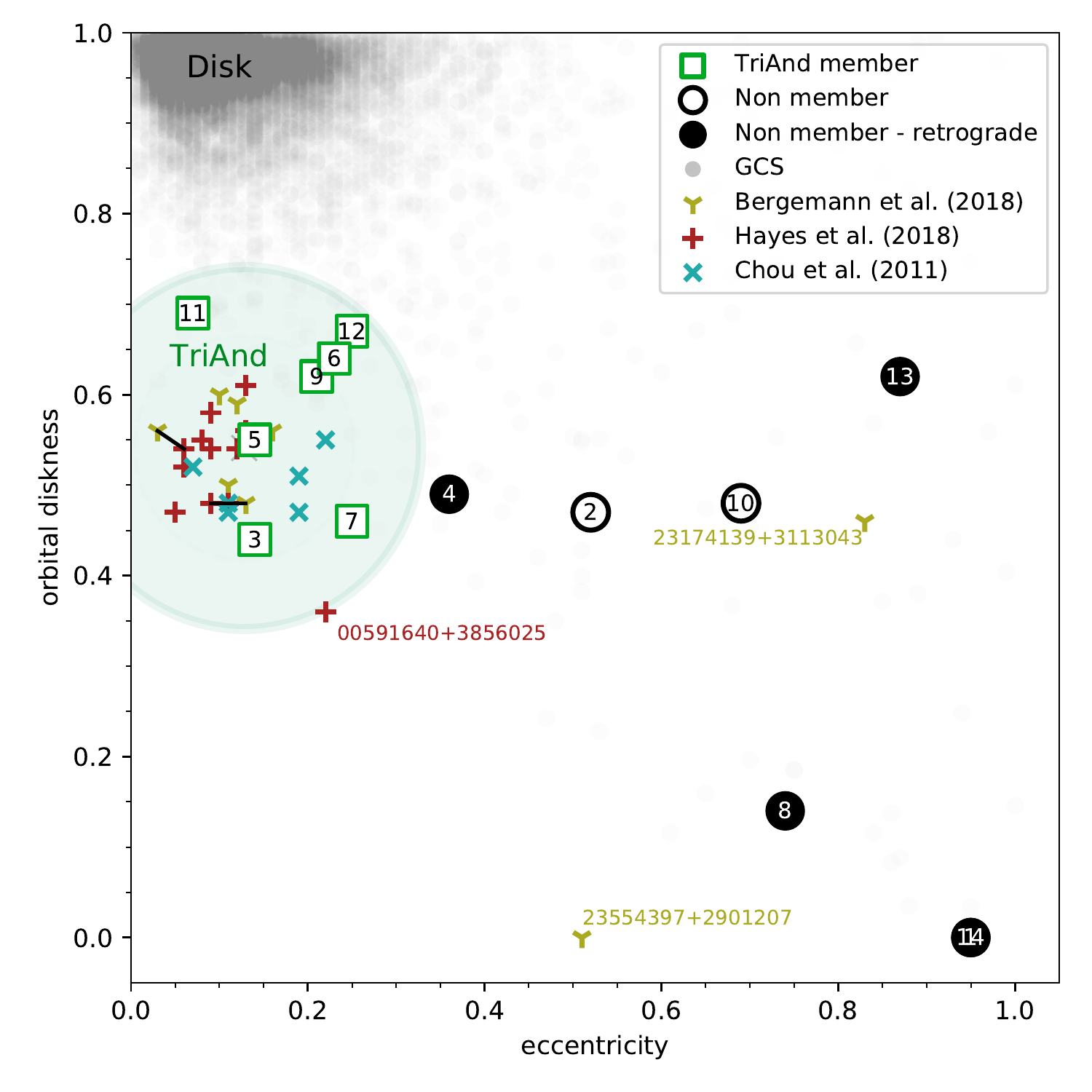}
\caption{Orbital characterization through eccentricity and orbital diskness (defined in Equations \ref{eq:ecc} and \ref{eq:od}) for the stars in our sample: the big green squares represent the stars classified as TriAnd members, while the big black circles represent the stars classified as non-members (filled black circles indicate that the star has a retrograde orbit). The stars in our sample are labeled according to the definition in Table \ref{table:observed_candidates}. The cyan exes correspond to the \citet{Chou2011} sample, the red crosses correspond to the \citet{Hayes2018} sample and the yellow ypsilon-like symbols to the \citet{Bergemann2018} sample of classified TriAnd stars. Excluding the labelled outliers, the stars from these two samples were used to find the centroid of TriAnd in this diagram. The green filled circle represents a radius of 0.2 in eccentricity and orbital diskness around the estimated centroid and were used to select the stars. For comparison, nearby disk stars from the Geneva-Copenhagen survey are shown in grey.}
\label{fig:ecc_disk}
\end{figure}

The characterized orbital parameters for the 14 stars in our sample are given in Table \ref{tab:orbital_parameters}. In Figure \ref{fig:ecc_disk}, we show the eccentricity and orbital diskness of our sample stars (big green squares and big black circles). The obtained orbital parameters for the stars from the samples in \citet{Chou2011}, \citet{Hayes2018}, and \citet{Bergemann2018} are included as well (cyan exes, red crosses, and yellow ypsilon-like symbols, respectively). To represent the location typically occupied by nearby disk stars in this diagram, we also include the Geneva-Copenhagen Survey stars (small gray circles; \citealp{Casagrande2011}). We see that, except for a few outliers, the stars from the \citet{Hayes2018} and \citet{Bergemann2018} occupy a very concentrated region of this diagram. We calculate the centroid of the distribution of these stars, excluding the three indicated outliers, and select all stars within a 0.2 radius around this centroid (green filled circle). Using this selection criteria, we classify as members the stars \#3, \#5, \#6, \#7, \#9, \#11 and \#12; star \#4 is not considered a member because it has a retrograde orbit and, therefore, is not kinematically similar to the other stars in the group. The stars selected by this criteria are exactly the same members selected using only the proper motions, but this has the advantage of taking into account the full 6D information regarding the stellar position and motion.

The diagram in Figure \ref{fig:ecc_disk} seems to be particularly robust in regard to the stellar distance, which is particularly difficult to estimate for stars that are located so far away from the Sun. The stars from  \citet{Bergemann2018} and \citet{Hayes2018} occupy the same location of the diagram even though the average of their reported stellar distances are very different among both samples: 13.0 and 20.0 kpc, respectively. Two stars are present in both samples: 2M00523040+3933030 and 2M01540851+3820287. They are indicated in Figure \ref{fig:ecc_disk} by the black lines connecting the red crosses and the yellow ypsilon-like symbols, and show that these orbital parameters do not change significantly for a change of a few kpc (3.0 and 3.5, kpc respectively). This result justifies our use of roughly estimated photometric parallaxes as a proxy of stellar distances, and allows us to consider the classification as very robust even though uncertainties in individual stellar distances are high.

\begin{table*}
\tabcolsep 0.15truecm
\small
\caption{The input astrometric data, distances and radial velocities used to integrate the stellar orbits using the python library \texttt{galpy} \citep{Bovy2015} and the obtained orbital parameters.}
\begin{tabular}{ll|cccccc|ccccc}\hline\hline
\multicolumn{13}{c}{TriAnd stars}\\\hline
\# &ID              &RA         &DEC        &dist  &pmra    &pmde    &rv      &$R_\mathrm{apo}$&$R_\mathrm{peri}$&$z_\mathrm{max}$ &ecc &disk- \\
   &                &hh mm ss.ss&dd mm ss.ss&kpc   &mas/yr  &mas/yr  &km/s    &kpc  &kpc  &kpc  &  & ness \\\hline
3 &00594094+4614332&00 59 40.95&46 14 33.23&23.921&$-0.151$&$ 0.064$&$-165.6$&30.12&22.63&11.64&0.14&0.44 \\
5 &01151944+4713512&01 15 19.45&47 13 51.23&20.370&$-0.294$&$-0.021$&$-162.4$&28.39&21.30& 8.16&0.14&0.55 \\
6 &02485891+4312154&02 48 58.91&43 12 15.44&16.255&$ 0.195$&$-0.199$&$-139.5$&26.23&16.28& 5.85&0.23&0.64 \\
7 &23535441+3449575&23 53 54.41&34 49 57.51&10.450&$-0.435$&$-0.721$&$-118.6$&14.99& 9.06& 5.49&0.25&0.46 \\
9 &02350813+4455263&02 35 08.14&44 55 26.30&27.802&$ 0.180$&$-0.174$&$-139.7$&35.96&23.61& 8.40&0.21&0.62 \\
11&02510349+4342045&02 51 03.50&43 42 04.54&20.508&$-0.061$&$-0.230$&$-106.2$&29.68&25.63& 5.42&0.07&0.69 \\
12&02475442+4429269&02 47 54.42&44 29 27.00&27.608&$ 0.166$&$-0.264$&$-145.2$&37.19&22.46& 7.34&0.25&0.67 \\\hline
\multicolumn{13}{c}{Non-TriAnd stars}\\\hline
1 &00075751+3359414&00 07 57.51&33 59 41.42&26.306&$ 0.827$&$-0.449$&$-233.6$&31.23& 0.78&30.95&0.95&0.00 \\
2 &00534976+4626089&00 53 49.77&46 26 09.00& 9.977&$-0.741$&$ 0.124$&$-245.3$&22.43& 7.03& 7.99&0.52&0.47 \\
4 &01020943+4643251&01 02 09.43&46 43 25.13&38.988&$ 1.881$&$-0.764$&$-144.0$&83.89& 39.9&28.57&0.36&0.49 \\
8 &23481637+3129372&23 48 16.38&31 29 37.21&27.074&$ 1.205$&$-0.992$&$-196.9$&30.23& 4.51&22.82&0.74&0.14 \\
10&23495808+3445569&23 49 58.08&34 45 56.92&10.667&$ 1.753$&$-0.488$&$ -43.4$&19.40& 3.56& 6.79&0.69&0.48 \\
13&02463235+4314481&02 46 32.36&43 14 48.15&25.194&$ 1.445$&$-1.224$&$-182.1$&33.50& 2.34& 7.94&0.87&0.62 \\
14&23174139+3113043&23 17 41.39&31 13 04.33&22.191&$ 1.294$&$-0.972$&$ -85.1$&30.15& 0.79&30.09&0.95&0.00 \\\hline
\hline
\end{tabular}
\label{tab:orbital_parameters}
\end{table*}

The stars \#1, \#2,  \#10, \#13 and \#14 are not compatible with TriAnd population according to three different criteria. Stars \#4 and \#8 are not compatible in proper motion and orbital parameters. Since these stars are not members in one or more criteria we considered them as non-members of the TriAnd overdensity. We also applied the three criteria at the \citet{Bergemann2018} and \citet{Hayes2018} samples. For the \citet{Chou2011} sample we applied only the proper motion criterion as the radial velocities of the stars in their sample were not estimated. Three stars from the \citet{Bergemann2018} sample --- 2M23484978+4549245, 2M23174139+3113043 (star \#14), and 2M23554397+2901207 ---  are not inside the ellipsoid of proper motion and two of them (2M23174139+3113043 and 2M23554397+2901207) do not have the orbital characteristic similar to the other TriAnd candidates (see Figure \ref{fig:ecc_disk}). 
Since the star 2M23484978+4549245 from \citet{Bergemann2018} could follow the proper motion of stars at that region (see Figure \ref{fig:proper_motion}) and it is inside of the selection region in Figure \ref{fig:ecc_disk}, we decided not to remove this star from the sample \citet{Bergemann2018}. 

\subsection{Chemical abundances}

\subsubsection{Metallicities and Iron-peak Elements}

Our investigation about the chemical nature of the TriAnd overdensity starts with the analysis of the metallicities and the abundances of the iron-peak elements Cr and Ni in the target stars. 
SNe type Ia is the main source of enrichment of the iron peak elements (Ni and Cr, as well as Fe) in the interstellar medium (\citealt{Iwamoto1999}).

\begin{figure}
\centering
\includegraphics[width=\columnwidth]{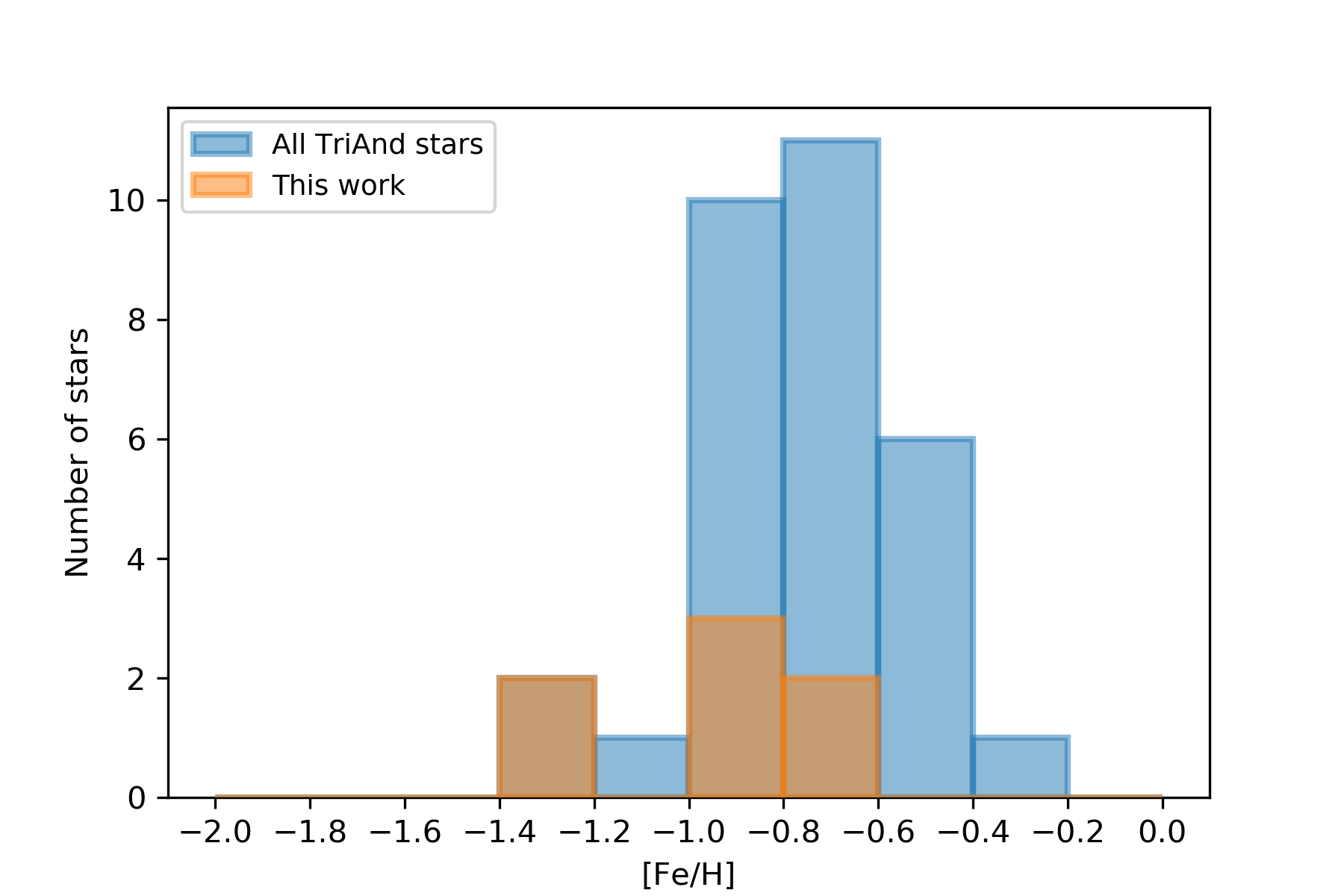}
\caption{Metallicity histogram for all TriAnd stars analyzed by high-resolution spectroscopy. The metallicity distribution for our TriAnd star sample is in orange.}%
\label{HistFeH}
\end{figure}

\begin{figure}
\centering
\includegraphics[width=\columnwidth]{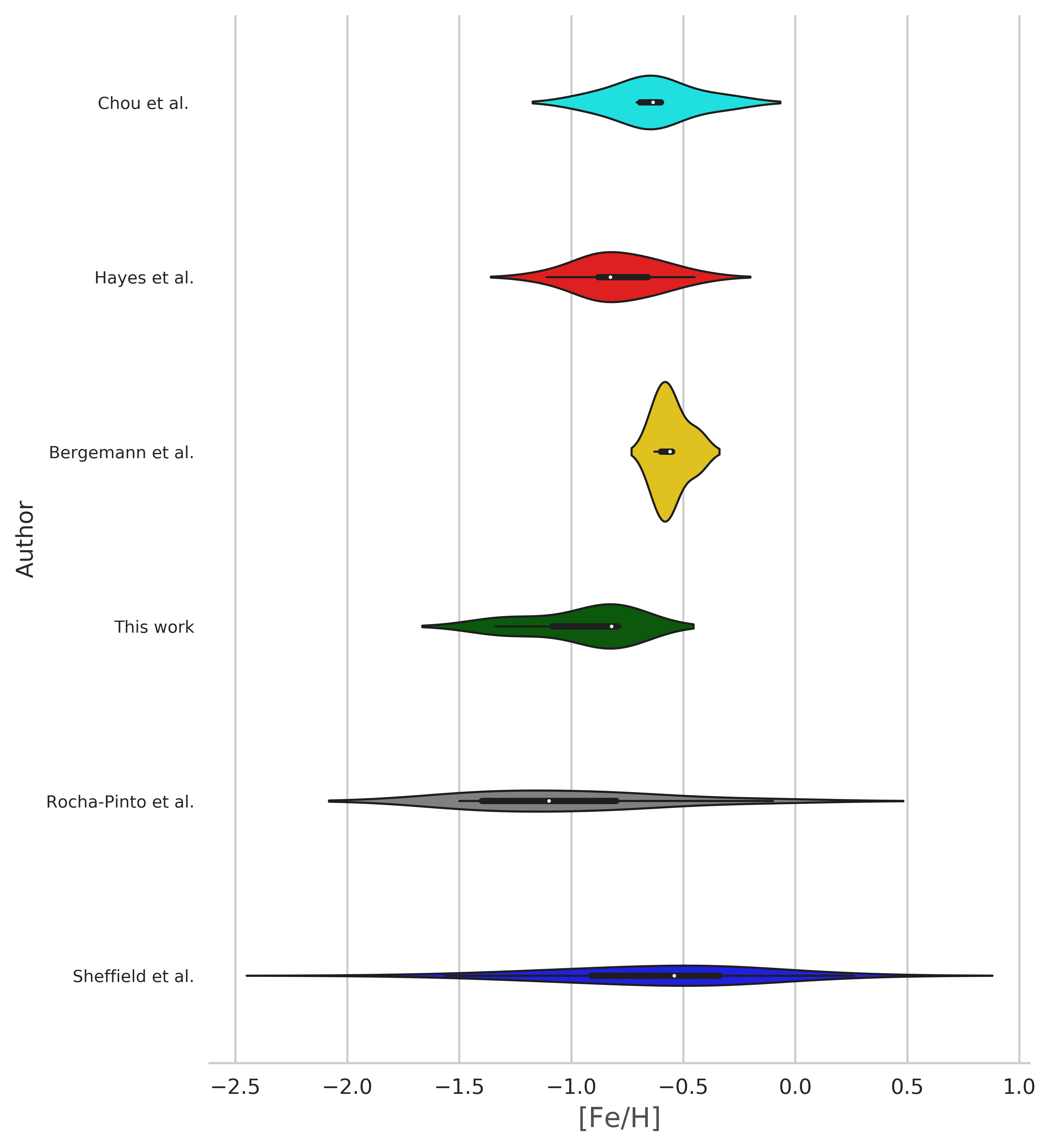}
\caption{Metallicity distribution for the TriAnd stars. In cyan: \citet{Chou2011}; in red: \citet{Hayes2018}; in yellow: \citet{Bergemann2018}; in green: this work; in grey: \citet{RP2004}, and in blue: \citet{Sheffield2014}. In violin plots, the white dot is the median, the thick bar is the interquartile range, and the thin bar represents the 95\% confidence interval. The distribution around these lines represents the distribution shape of the sample where the wider regions represent a higher probability that a star will have that metallicity.}%
\label{fig:FeH}
\end{figure}

The TriAnd kinematically confirmed stars span the metallicity range between $-$1.34 $\pm$ 0.12 $\leq$ [Fe/H] $\leq -$0.78 $\pm$ 0.1 dex (see Table \ref{table:parameters}), distributed in two metalicity groups, one more metal-rich group (five TriAnd stars) with $-$0.94 $\pm$ 0.1 $\leq$ [Fe/H] $\leq -$0.78 $\pm$ 0.1 and one metal-poor group (two TriAnd stars) with $-$1.34 $\pm$ 0.12 $\leq$ [Fe/H] $\leq -$1.23 $\pm$ 0.09 (see Figure \ref{HistFeH}); the possible distribution in two metallicity groups, however, is not considered to be significant given the small number of stars in our sample. 
In Figure \ref{fig:FeH} we show the metallicity distribution of the members compared to the distributions obtained in other studies for TriAnd.  The metallicity distribution for the TriAnd stars of our sample (shown in green) generally agrees with the metallicity distribution of \citet{Hayes2018} (shown in red), although our distribution has a metal-poor metallicity tail. The metallicity distribution of \citet{Bergemann2018}'s sample is on average more metal-rich, in rough agreement with the average from \citet{Chou2011}, although \citet{Bergemann2018} results show less scatter.
It is expected that some of the differences in the metallicity results shown in Figure \ref{HistFeH} are due, in part, to the different methodologies adopted in the different studies: different line lists, different model atmospheres, LTE versus non-LTE, optical versus infrared. For example, \citet{Bergemann2018} determined a metallicity of 0.24 dex larger than that obtained in APOGEE DR14 (\citealt{Hayes2018}) for the star 2M00523040+3933030, while for another star both sets of results agree. 
For the star \#14 we derived a very similar metallicity compared to that obtained by \citet{Bergemann2018}. In Figure \ref{fig:FeH} we present the metallicity distribution for all TriAnd stars analyzed with high resolution spectroscopy (our study, \citealt{Chou2011}, \citealt{Hayes2018}, \citealt{Bergemann2018}) and confirmed as TriAnd members through the criteria shown in subsection \ref{sec:kinematics}. The metallicity distribution for all TriAnd stars is characterized by a peak between -0.6 and -1 dex. 

Interestingly, two TriAnd stars in our sample have slightly lower metallicities (star \#12 with [Fe/H] $-$1.34 $\pm$ 0.12 and star \#9 with [Fe/H] $-$1.23 $\pm$ 0.09) than those found in previous high-resolution spectroscopic studies in the literature. The metallicities of the TriAnd stars obtained in this study are, in fact, very similar to the results for TriAnd from \citet{Deason2014} obtained using low-resolution spectra from Sloan Extension for Galactic Understanding and Exploration (SEGUE). \citet{Deason2014} found a metallicity range for the TriAnd stars from $-0.5$ to $-1.3$ dex. We note that \citet{Hayes2018} also has one more metal-poor star in the APOGEE TriAnd sample ([Fe/H]= $-$1.1).

\begin{table*} 
\tabcolsep 0.32truecm
\caption{Abundance ratios([X/Fe]) for the observed stars.}
\begin{tabular}{llccccccc}\hline\hline
\multicolumn{9}{c}{TriAnd stars}\\\hline
\# & ID               &     [Na/Fe]    &  [Al/Fe]       &[Cr/Fe] &   [Ni/Fe]     &[Ba/Fe] & [Eu/Fe] &   [Eu/Ba]\\\hline        
3 & 00594094+4614332 &  0.34$\pm$0.09 &  0.36$\pm$0.05 & ---    & $-$0.02$\pm$0.12&  0.11  &   0.09  &   $-$0.02  \\
5 & 01151944+4713512 &  0.36$\pm$0.12 &  0.38$\pm$0.05 & ---    &  0.16$\pm$0.12&  0.21  &   0.26  &    0.05  \\
6 & 02485891+4312154 &  0.16$\pm$0.14 &  0.39$\pm$0.07 & ---    &  0.06$\pm$0.11&  0.37  &   0.28  &   $-$0.09  \\
7 & 23535441+3449575 & $-$0.23$\pm$0.09 &  0.10$\pm$0.14 &$-$0.01   & $-$0.03$\pm$0.09&  0.46  &   0.40  &   $-$0.06  \\
9 & 02350813+4455263 &  0.17$\pm$0.13 &  0.51$\pm$0.08 & 0.08   &  0.09$\pm$0.13&  0.22  &   0.15  &   $-$0.07  \\
11 & 02510349+4342045 &  0.15$\pm$0.03 &  0.24$\pm$0.13 &$-$0.21   &  0.14$\pm$0.12&  0.58  &   0.30  &   $-$0.28  \\
12 & 02475442+4429269 &  0.26$\pm$0.13 &  0.53$\pm$0.09 &$-$0.07   &  0.25$\pm$0.13&  $-$0.11 &   0.26  &    0.37  \\\hline
\multicolumn{9}{c}{Non-TriAnd stars}\\\hline
1 & 00075751+3359414 & $-$0.14$\pm$0.04 &  0.26$\pm$0.13 & $-$0.13  & $-$0.05$\pm$0.13&  0.21  &   0.71  &    0.50  \\
2 & 00534976+4626089 & $-$0.09$\pm$0.03 &  0.12$\pm$0.13 & ---    &  0.01$\pm$0.13&  0.14  &   0.28  &    0.14  \\
4 & 01020943+4643251 & $-$0.03$\pm$0.08 &  0.15$\pm$0.11 & $-$0.52  &  0.06$\pm$0.11&   0.05 &   0.57  &    0.52  \\
8 & 23481637+3129372 & $-$0.13$\pm$0.09 &      ---       & $-$0.23  &  0.02$\pm$0.12&  0.17  &   0.84  &    0.67  \\
10 & 23495808+3445569 & $-$0.50$\pm$0.09 & $-$0.31$\pm$0.12 & ---    & $-$0.03$\pm$0.12&  0.40  &   0.70  &    0.30  \\
13 & 02463235+4314481 &  0.08$\pm$0.13 &  0.25$\pm$0.03 &$-$0.16   & $-$0.01$\pm$0.14&  0.22  &   0.53  &    0.31  \\
14 & 23174139+3113043 & $-$0.31          &  0.04          &$-$0.14   & $-$0.17$\pm$0.14&  0.42  &   0.70  &    0.28  \\\hline
\hline
\end{tabular}
\label{tab:abundance_ratio}
\end{table*}

\begin{figure*}
\centering
\includegraphics[width=15cm]{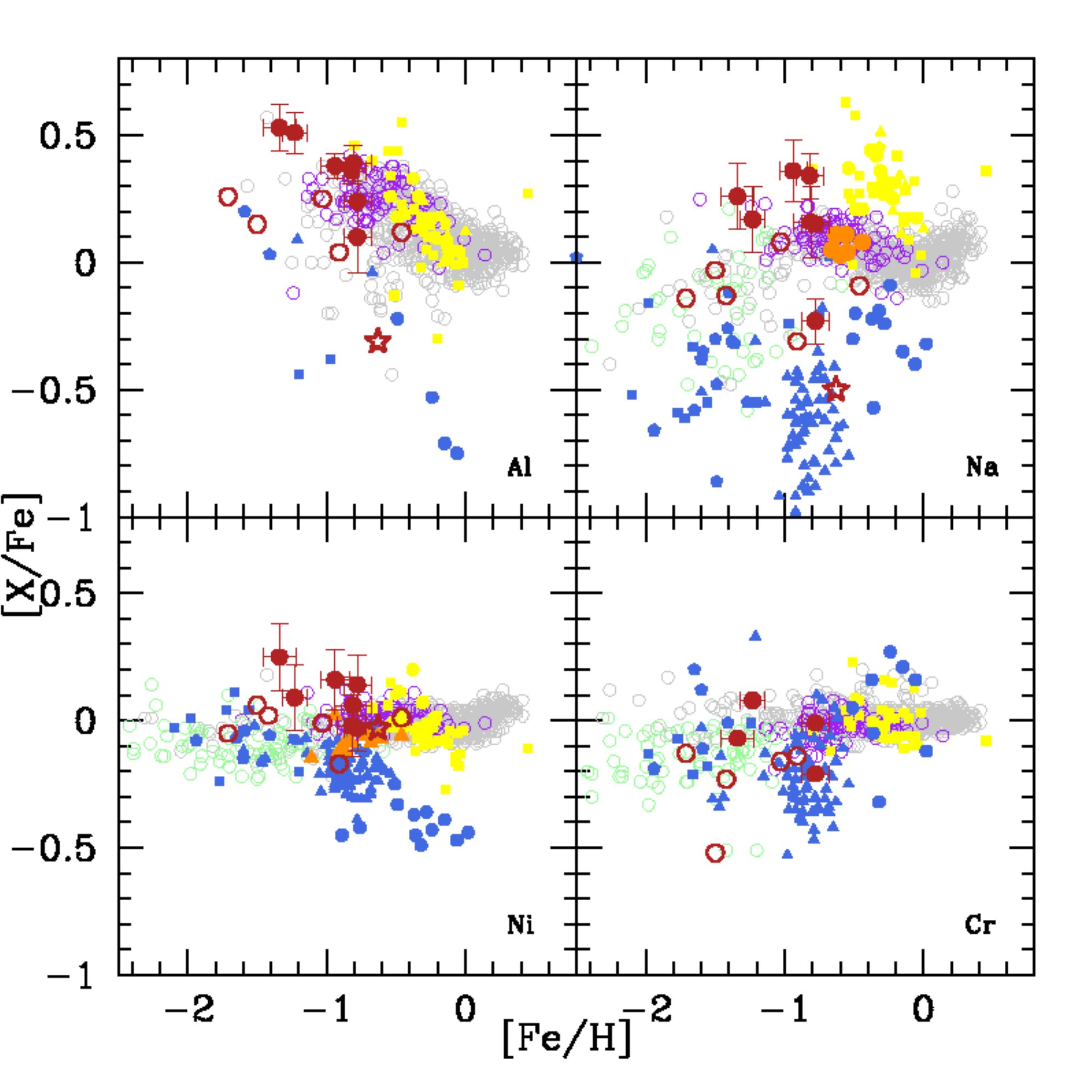}
\caption{Abundance ratios [X/Fe] vs. [Fe/H]. 
Red circles: our TriAnd stars sample; 
Open red symbols: our non-TriAnd sample; 
Orange triangles: TriAnd stars from \citet{Hayes2018}; 
Orange circles: TriAnd stars from \citet{Bergemann2018}; 
Gray circles: local disk stars from \citet{Bensby2014}; 
Purple circles: thick disk stars from \citet{Reddy2006}; 
Yellow circles: open clusters from the outer disk from \citet{Yong2012}; Yellow squares: Cepheids from outer disk from \citet{Luck2011}; Yellow hexagons: Cepheids from outer disk from \citet{Lemasle2013}; Yellow triangles: Cepheids from outer disk from \citet{Genovali2015}; Green circles: halo stars from \citet{Ishigaki2012}; 
Blue squares: stars from Sculptor dwarf Galaxy from
\citet{Geisler2005} and \citet{Shetrone2003}; 
Blue hexagons: stars from Carina dwarf galaxy from \citet{Koch2008}, and \citet{Shetrone2003}; 
Blue triangles: stars from Fornax dwarf galaxy from \citet{Shetrone2003} and \citet{Letarte2010}; 
Blue circles: stars from Sagittarius dwarf galaxy from \citet{Sbordone2007}. 
}%
\label{fig:alFieldartigo3final}
\end{figure*}

The TriAnd stars in our sample exhibit [Ni/Fe] in the range between $-$0.03 $\pm$ 0.09 $\leq$ [Ni/Fe] $\leq$ $+$0.25 $\pm$ 0.13 and [Cr/Fe] between $-0.21$ to +0.08 dex (Table \ref{tab:abundance_ratio}); this is the first study to obtain chromium abundances in TriAnd stars.
In Figure \ref{fig:alFieldartigo3final} the results for [Cr/Fe] and [Ni/Fe] are shown as a function of [Fe/H] for our sample (filled red circles). 
We also show in this figure our results for non-TriAnd stars (as open red symbols), as well as literature results for the TriAnd stars (\citealt{Chou2011}, \citealt{Bergemann2018}, \citealt{Hayes2018}), local disk stars (\citealt{Bensby2014}), thick disk stars (\citealt{Reddy2006}), open clusters from the outer disk (\citealt{Yong2012}), cepheids from the outer disk (\citealt{Luck2011}, \citealt{Lemasle2013} and \citealt{Genovali2015}), halo stars (\citealt{Ishigaki2012}), and stars from dwarf galaxies (Sculptor: \citealt{Shetrone2003}, \citealt{Geisler2005}; Carina: \citealt{Shetrone2003}, \citealt{Koch2008}; Fornax: \citealt{Shetrone2003}, \citealt{Letarte2010}; Sagittarius: \citealt{Monaco2005}, \citealt{Sbordone2007}). 
The majority of our TriAnd sample presents [Cr/Fe] and [Ni/Fe] ratios similar to the chemical pattern of the local disk. This is case in particular for Cr. However, for Ni, the lower metallicity TriAnd stars in our sample tend to show a slight overabundance of Ni, which is not seen in the results for the local disk stars. The APOGEE results (orange triangles) do not probe such low metallicities.

Simulations indicate that Ni and Fe yields in SNe type Ia are strongly dependent on the white dwarf mass (\citealt{LeungNomoto2018}). Thus, the slight overabundance of the [Ni/Fe] ratio in the TriAnd stars of lower metallicity in relation to the local disk may be due to the mass difference of the white dwarfs that produced Ni and Fe in the TriAnd region and the local disk. In Figure \ref{fig:alFieldartigo3final} we see that the Sagittarius stars (blue circles) present a sub-solar [Ni/Fe] ratio at [Fe/H] $= -0.8$, a chemical pattern not observed for the TriAnd stars.

\subsubsection{Na and Al}

Na and Al are mainly formed during the evolution of massive stars (\citealt{WoosleyWeaver1995}). Na abundances in the surface of red giants is also affected by mixing processes occurring in the stellar interior (\citealt{CharbonnelLagarde2010}).

The [Na/Fe] ratio obtained for the TriAnd stars studied here spans the range from $-$0.23 $\pm$ 0.09 to $+$0.36 $\pm$ 0.12 dex, with all stars, except one, exhibiting [Na/Fe] ratios greater than zero (Figure \ref{fig:alFieldartigo3final}). Most of the TriAnd stars in our sample (filled red circles) show Na overabundances when compared to the local disk stars in the same metallicity range (Figure \ref{fig:alFieldartigo3final}), with some overlap with the thick disk results from Reddy et al. (2006). 
An overabundance in [Na/Fe] is also observed for the open clusters and cepheids in the outer disk (shown as yellow circles in Figure \ref{fig:alFieldartigo3final}), although these have higher metallicities than our TriAnd sample; in addition, we note that the outer disk clusters and cepheids lie at much closer distances than the TriAnd overdensity, not probing the same region of the Galactic disk. The results from  \citet{Bergemann2018} sample present a typical [Na/Fe] ratio of local disk stars, such results are not necessarily inconsistent with ours, as their sample probes a higher metallicity regime for TriAnd. 
In addition, the non-TriAnd stars in our sample have lower [Na/Fe] with a larger scatter with a behavior that is clearly distinct of the TriAnd stars.

One of our TriAnd star (star \#7), however, has a much lower [Na/Fe] ratio ([Na/Fe] = $-$0.23 $\pm$ 0.09), its chemistry would indicate that it could possibly be from a different population, the [Na/Fe] ratio for this star lies at the upper envelope of the dwarf spheroidal results for [Na/Fe] (presented as blue symbols in Figure \ref{fig:alFieldartigo3final}), which are much lower than the local disk. We note that such low values for [Na/Fe] are also found for two additional target stars (star \#10 and \#14) that have not been kinematically confirmed here as part of the TriAnd overdensity.  
\citet{Bergemann2018} classified star \#14 (for which we find [Na/Fe] = $-0.31$ and \citet{Bergemann2018} find [Na/Fe] = -0.30) as being from the TriAnd overdensity. However, as discussed previously, this star has a proper motion that is different from that of the other TriAnd stars, suggesting that it does not belong to TriAnd. 
The non-TriAnd star \#10, having the lowest [Na/Fe] ratio in our sample (= $-$0.5 $\pm$ 0.09 dex; shown as the red star symbol in Figure \ref{fig:alFieldartigo3final}), exhibits a chemical pattern in many elements similar to that of dwarf galaxy stars, showing that the field in the direction of TriAnd overdensity is composed of a stellar population mixture. 

Adopting the kinematical definition for TriAnd stars in this study, the [Na/Fe] results obtained would indicate that the TriAnd overdensity presents a large scatter in [Na/Fe] and multiple chemical patterns, with high [Na/Fe] for most stars, but with one TriAnd star having much lower [Na/Fe], which is consistent with the chemical pattern of dwarf spheroidal galaxies.  Despite this low [Na/Fe] result, our results taken in conjunction with those from \citet{Bergemann2018}, may indicate that the [Na/Fe] ratio in the TriAnd population may increase slightly with the decrease in metallicity (Figure \ref{fig:alFieldartigo3final}). (See also discussion in \citet{Smiljanic2016} that the chemical evolution of Na in the Galaxy is not well understood).
 
This is the first study to present aluminum abundances for TriAnd members. We find that the TriAnd stars have overabundance of Al with respect to Fe, with [Al/Fe] ranging from $+$0.10 $\pm$ 0.14 to $+$0.53 $\pm$ 0.09 dex (Table \ref{tab:abundance_ratio}). The [Al/Fe] pattern in the TriAnd stars mostly overlaps with that of the thick disk stars from Reddy et al. (2006; shown as open purple symbols in Figure \ref{fig:alFieldartigo3final}). However, the lowest metalicity stars in our sample show significantly higher values of [Al/Fe] than the decreasing [Al/Fe] results for [Fe/H] $< -1$ dex obtained in \citealt{Bensby2014} and \citealt{Ishigaki2012}; the behavior of TriAnd at low metallicity seems to increase for the lowest metallicity stars probed, while it is important to note that the non-TriAnd stars in our sample show an overall a distinct behavior at low metallicity. Very few results are shown in figure \ref{fig:alFieldartigo3final} for dwarf spheroidals; similarly to Na, it is clear that the non-TriAnd star \#10 shows a low [Al/Fe] value, which is consistent with dwarf spheroidals. 

\subsubsection{Neutron-capture Elements}

The s-process elements are mainly produced during the Asymptotic Giant Branch phase (\citealt{Busso1999}), whereas the r-process elements are believed to be mainly produced in merging neutron stars (\citealt{Thielemann2017}) and possibly also in the explosive phase of type II supernovae (\citealt{Thielemann2002}). In general, both s- and r- processes contribute to the production of elements heavier than Fe. In this study, we derive barium and europium abundances for the target stars, to gauge the respective contributions of the s- and r- processes in the TriAnd stars.

\begin{figure*}
\centering
\includegraphics[width=15cm]{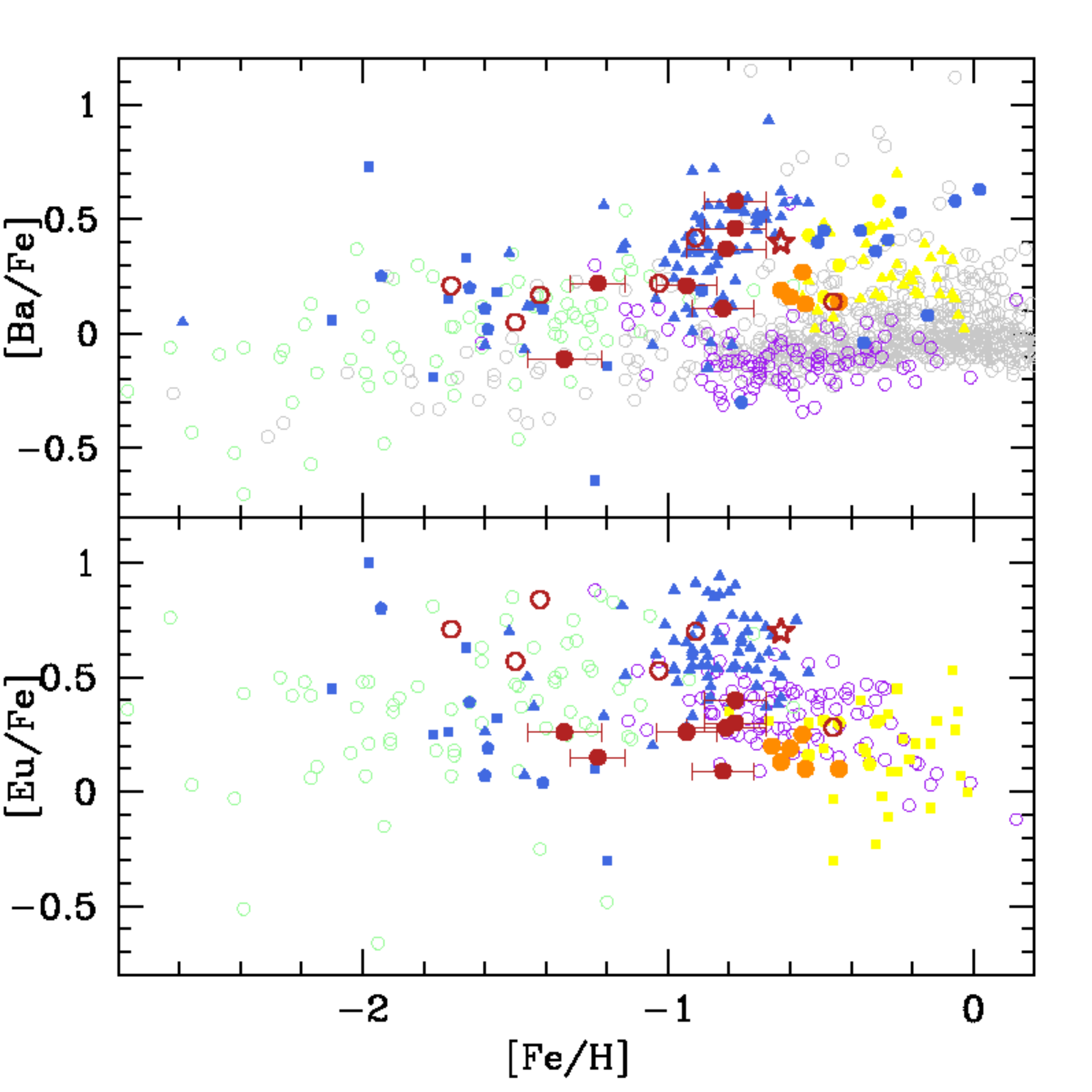}
\caption{Abundance ratios [X/Fe] vs. [Fe/H]. Symbols have the same meaning as in Figure \ref{fig:alFieldartigo3final}. Yellow triangles in upper panel: Cepheids from outer disk from \citet{Andrievsky2014}.}%
\label{fig:BaEudwarfgalaxiesartigofinal}
\end{figure*}

Table \ref{tab:abundance_ratio} presents the barium and europium results obtained for all target stars. 
In Figure \ref{fig:BaEudwarfgalaxiesartigofinal} we show the derived [Ba/Fe] and [Eu/Fe] ratios along with the results from  \citet{Bergemann2018}, local disk (\citealt{Bensby2014}), open clusters and cepheids of the outer disk (\citealt{Luck2011}, \citealt{Yong2012}, \citealt{Lemasle2013} and \citealt{Genovali2015}), and dwarf galaxies. 
The [Ba/Fe] ratio for the TriAnd stars presents a large dispersion, with a range between $-0.11$ - +0.58 dex. Our results would seem to indicate that [Ba/Fe] decreases with decreasing metallicity. In general, the TriAnd stars show a Ba overabundance with respect to Fe, with only one star (\#12) showing a negative value of [Ba/Fe] = $-0.11$. 
Our [Ba/Fe] results mostly overlap with those from dwarf spheroidals (in particular for metallicities higher than $\sim$ -1.0 dex).  The results for \citet{Bergemann2018} sample of TriAnd stars (orange circles) indicate a [Ba/Fe] ratio $\sim$ 0.18 $\pm$ 0.06 dex, being slightly higher than that for most of the local disk stars by \citet{Bensby2014} but still falling below the dwarf spheroidal results shown in the figure. 

The [Eu/Fe] ratios for all TriAnd stars in our study span the range between +0.09 to +0.40 dex, exhibiting an approximately constant trend with metallicity. The results from \citet{Bergemann2018} (orange circles) would seem to extend the roughly constant [Eu/Fe] behavior towards higher metallicities. It is interesting to note that those stars in our sample that are not from TriAnd (open red circles) show a very different behavior when compared to TriAnd, including the star \#10 that shows again an abundance pattern in line with those of dwarf spheroidals, in this case Fornax, having a high value of [Eu/Fe].
To summarize, the [Eu/Fe] ratios are overall lower than results from dwarf spheroidals, showing a distinct behavior when compared to the s-process element barium that follows the pattern observed for Fornax.

\begin{figure}
\centering
\includegraphics[width=\columnwidth]{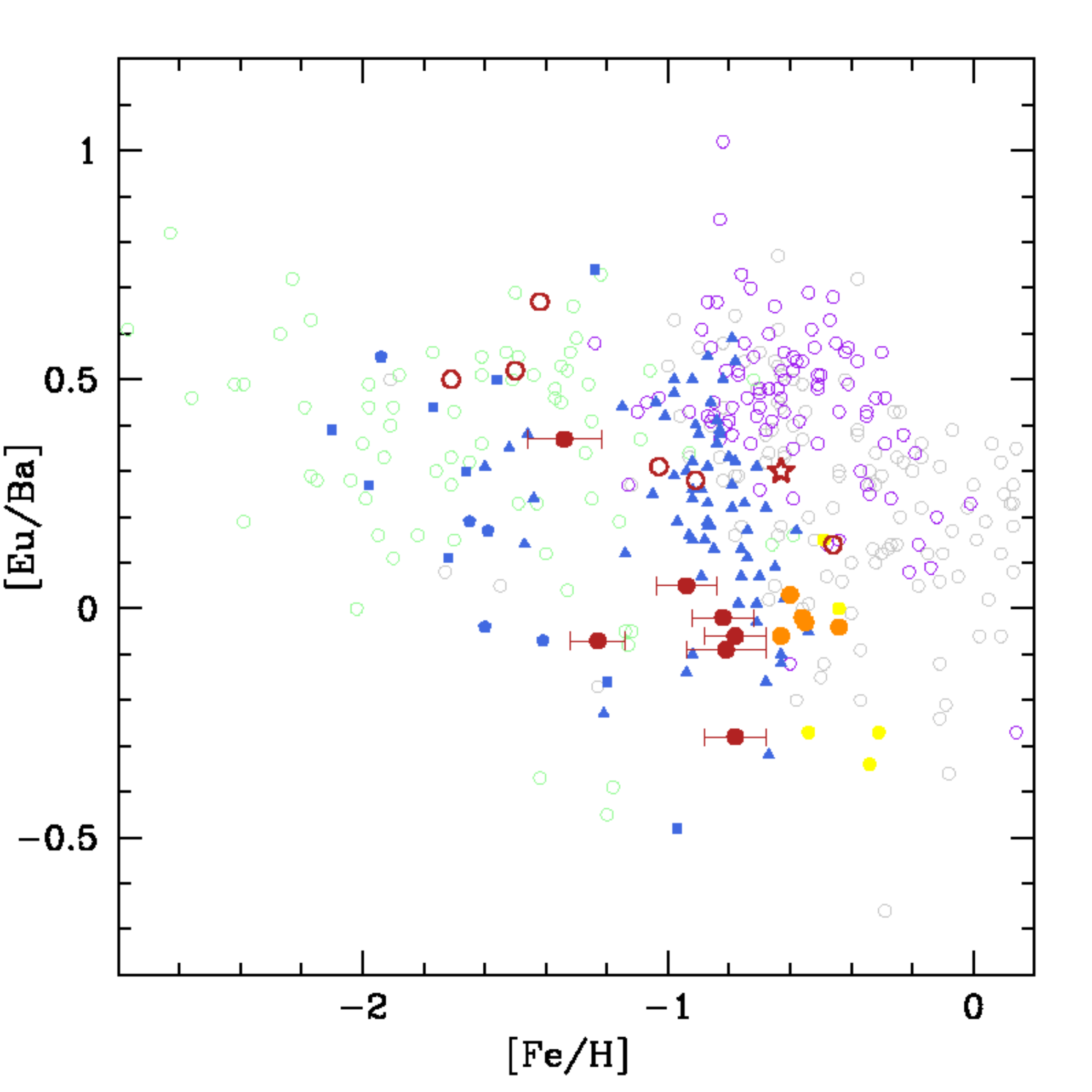}
\caption{Abundance ratios [Ba/Eu] vs. [Fe/H]. Symbols have the same meaning as in Figure \ref{fig:alFieldartigo3final}. The open gray circles represent the local disk stars from \citet{Edvardsson1993}, \citet{NissenSchuster1997}, \citet{Hanson1998}, \citet{Prochaska2000}, \citet{Fulbright2000}, \citet{Fulbright2002}, \citet{StephensBoesgaard2002}, \citet{Bensby2003}, and \citet{Reddy2003}, as compiled by \citet{Venn2004}.}%
\label{fig:EuBaFeHartigofinal}
\end{figure}

The [Eu/Ba] abundance ratio, a monitor of the r- and s- processes contribution in the interstellar medium that formed the TriAnd stars, is presented in Figure \ref{fig:EuBaFeHartigofinal}. This clearly shows how TriAnd stars segregate when compared to the disk stars. Here again the results from \citet{Bergemann2018} and from this study indicate a constant pattern for the [Eu/Ba] ratios. The lowest metallicity star in our TriAnd sample (star \#12) has a high [Eu/Ba] ratio ([Ba/Fe]= $-0.11$ dex and [Eu/Fe]= 0.26 dex). Taken a face value this could indicate a higher value for [Eu/Ba] for the lowest metallicities in TriAnd, but it should be kept in mind this is found for only one star in our TriAnd sample. 
The different [Eu/Ba] ratios in the TriAnd stars may be evidence of a distinct abundance pattern for stars belonging to the TriAnd overdensity region.

\subsubsection{Abundance comparisons}

In order to check for possible systematic differences with the results presented in \citet{Bergemann2018} when compared to ours, we analyzed one star (\#14) in common with that study and find excellent agreement in the metallicity with a difference: $\Delta$ [Fe/H]=0.02. 
The differences in the abundances derived for the other elements in common in the two studies are also small: $\Delta$ [Na/Fe]=0.02, $\Delta$ [Ba/Fe]=0.11, $\Delta$ [Eu/Fe]=0.02. 
This comparison, although for only one star, indicates similar abundance results in the two studies without any important systematic abundance differences.

\section{Discussion}

The galactic nature of the TriAnd overdensity began to be unraveled through the high-resolution spectroscopic studies by \citealt{Bergemann2018}, and \citealt{Hayes2018}. 
The interactions of our Galaxy with neighboring galaxies could, for example, induce a warped and flared disk, or vertical density waves, and explain overdensities, such as TriAnd (e.g., \citealt{Momany2006}, \citealt{Gomez2013}, \citealt{Bergemann2018}).
\citet{Bergemann2018} found that TriAnd stars have an abundance pattern similar to that of Galactic disk. 
The kinematic analysis presented here corroborates with the Galactic disk pattern of the TriAnd stars orbits.

The abundance results in this study indicate that the TriAnd population shows a complex chemical pattern exhibiting clear differences when compared to the local disk pattern (\citealt{Bensby2014}), such as, for example, for the [Na/Fe] and [Ba/Fe] ratios (see Figure \ref{fig:alFieldartigo3final}, and \ref{fig:BaEudwarfgalaxiesartigofinal}). 

In the following we summarize the detailed chemical pattern obtained for the sample of TriAnd stars studied here: 
(1) An extended metallicity distribution ranging between [Fe/H] $\sim$ $-$1.34 $\pm$ 0.12 to $-$0.78 $\pm$ 0.1 dex, which includes more metal-poor stars than the previous high-resolution studies from \citet{Chou2011}, \citet{Bergemann2018}, and \citet{Hayes2018}. 
(2) The [Na/Fe] ratios obtained for the TriAnd members are, for the most part, higher than the local disk. When combined with the [Na/Fe] results from \citet{Bergemann2018} for higher metallicity TriAnd stars, there is an indication that [Na/Fe] in the TriAnd population may increase with decreasing metallicity (Figure \ref{fig:alFieldartigo3final}). 
The presence of a confirmed TriAnd star having a low [Na/Fe] ratio, more similar to what is found for dwarf spheroidals, might perhaps indicate that TriAnd may have multiple populations, but this is a very speculative idea at this point.
(3) The [Al/Fe] ratios for TriAnd stars also exhibit an increase with the decrease in metallicity, similarly to what is observed for the local disk (\citealt{Bensby2014}). The two lowest metallicity TriAnd stars are an extension of the growth in the [Al/Fe] ratio for [Fe/H] $< - 1$.
(4) The [Ni/Fe] ratios for our sample of TriAnd stars show possible overabundances when compared to the local disk; for stars in the same metallicity range as the APOGEE sample in Hayes et al. (2018), our results are $\sim$ 0.1 dex higher than the APOGEE [Ni/Fe]. The differences with the local disk pattern may not be significant.
(5) In general, the barium in the TriAnd stars is overabundant and [Ba/Fe] ratios decrease with decreasing metallicity, showing a pattern similar to that presented by dwarf spheroidals.
(6) The [Eu/Fe] ratios for the TriAnd stars are approximately constant with metallicity (Figure \ref{fig:EuBaFeHartigofinal}), this result is in line with the [Eu/Fe] from \citet{Bergemann2018} for higher metallicities.
(7) The [Eu/Ba] ratio for the TriAnd stars in our sample have similar values those obtained in \citet{Bergemann2018} for higher metallicity TriAnd stars, indicating a roughly constant [Eu/Ba] ratio with metallicity. 
This abundance pattern is not seen in local disk stars, with TriAnd stars showing, in general, low values when compared to the other galactic populations. Such results indicate the predominance of enrichment in s-process elements when compared to r-process elements in the gas that formed TriAnd. The high value for [Eu/Ba] for the most metal-poor TriAnd star in our sample may indicate that [Eu/Ba] raises with decreasing metallicity.

The metallicity distribution for all TriAnd star samples obtained using high-resolution spectroscopy (our study, \citealt{Chou2011}, \citealt{Bergemann2018}, and \citealt{Hayes2018}) indicates a large dispersion, having one star with a metalicities as low as $\sim$ $-$1.34 $\pm$ 0.12 dex (Figure \ref{fig:FeH}).
Given the large galactocentric distance to TriAnd (R$_{GC} \gtrsim$ 20 kpc), it is of interest to compare its chemical pattern with that of the most distant stars known in the galactic disk, keeping in mind that overall the chemical pattern of the disk at galactocentric distances comparable to TriAnd (R$_{GC} \gtrsim$ 20 kpc) is still unknown.

For the outer disk regions (12 $\lesssim$ R$_{GC} \lesssim$ 20 kpc), high-resolution spectroscopic studies of Cepheids, open clusters and field red giants (\citealt{Andrievsky2002}, \citealt{Luck2003}, \citealt{Carraro2004}, \citealt{Yong2005}, \citealt{Carney2005}, \citealt{Yong2006}, \citealt{Bensby2011}, \citealt{Yong2012}, \citealt{Lemasle2013}, \citealt{Hayden2015} \citealt{CantatGaudin2016}, \citealt{Reddy2016}, \citealt{Carraro2017}, \citealt{Magrini2017}) show that the metallicities of these populations cover a range roughly between $-0.8$ $\leq$ [Fe/H] $\leq$ $-0.2$ dex; this is more metal rich than the metallicity range exhibited by the TriAnd stars in our sample ($-$1.34 $\pm$ 0.12 $\leq$ [Fe/H] $\leq -$0.78 $\pm$ 0.1) and most of APOGEE TriAnd sample discussed in \citet{Hayes2018} (Figure \ref{fig:FeH}). 
Considering the metallicities shown in Figure \ref{fig:FeH} for the combined TriAnd samples and the discussion from \citet{Hayes2018} of the metallicities and chemical abundances for field red giants located at different galactocentric distances in the APOGEE survey, we find that the mean metallicity for the TriAnd stars is overall consistent with the extrapolated abundance gradients in the outer disk.

TriAnd stars present high [Na/Fe] and [Ba/Fe] ratios when compared with the local disk (Figure \ref{fig:alFieldartigo3final}, and \ref{fig:BaEudwarfgalaxiesartigofinal}).
Similarly what is found for the TriAnd stars, the Cepheids and open clusters of the outer disk (12 kpc $\lesssim$ R$_{GC} \lesssim$ 20 kpc) also present [Na/Fe] and [Ba/Fe] greater than the local disk, although the Cepheids and open clusters are more metal-rich than TriAnd.
In addition, concerning the patterns for the elements Al, Na, Ni, Ba and Eu, TriAnd could be considered as an extension to lower metallicities of the abundance trends [X/Fe] ratios of outer disk between 12 kpc $\lesssim$ R$_{GC} \lesssim$ 20 kpc.
The [Ba/Fe] ratio of the TriAnd stars also resembles the pattern presented by the Fornax dwarf galaxy, however, we observe significant differences between these populations for other elements (like Na). We detected one Non-TriAnd star with a chemical abundance that resembles dwarf galaxies in the direction of the TriAnd overdensity (see the star symbol referring to the Non-TriAnd star 10 in Figures \ref{fig:alFieldartigo3final}, \ref{fig:BaEudwarfgalaxiesartigofinal} and \ref{fig:EuBaFeHartigofinal}) indicate that such regions may suffer from pollution of dwarf galaxies as also found by \citet{Chou2011}. 

Modeling the chemical evolution of the outermost regions of the Galactic disk is challenging given, on the one hand, the variety of mechanisms that influence the evolution of the outer Galaxy, and, on the other, the few high-resolution spectroscopic observations of stars in this region to constrain the models. The high uncertainty associated with the age and distance of the stars, even currently in the era of GAIA astrometric data, is a further complication. 

One of the most promising hypotheses to explain the TriAnd overdensity would be disk disruption caused by the interaction of the Galaxy with its neighbors (\citealt{Bergemann2018}), similar interactions were found in other spiral galaxies, such as ESO 510-G13 and NGC 1512 (\citealt{LopezSanchez2015}). In addition to interactions and mergers between galaxies, other mechanisms could perhaps be invoked to contribute to the chemical abundance patterns of outer disks, such as, stellar radial migration (\citealt{SellwoodBinney2002}), galactic winds (e.g., \citealt{Zhang2018}), as well as possible differences in the Initial Mass Function (IMF) and star formation rate, due to regions of lower gas densities in the outer disk when compared to the innermost disk regions.
The abundance patterns of the stars in the TriAnd overdensity obtained in this study, which is somewhat unique when compared to any other galactic population at its metallicity, highlight this complexity, indicating that many variables need to be considered for modeling this region. 
The next step in further understanding the nature TriAnd overdensity is the reliable homogeneous analysis of a much larger sample of this population; as well as other populations of the outer Milky Way disk, at R$_{GC} ~$ 20 kpc and beyond; the outer disk remains an unexplored territory that starts to be unraveled.

   \section{Conclusions}

The Triangulum--Andromeda (TriAnd) overdensity is a distant structure of the Milky Way (R$_{GC}$ $\sim$ 20 -- 30 kpc) located in the second Galactic quadrant well below the Galactic plane (\citealt{RP2004}; \citealt{Majewski2004}). 

We analysed high-resolution optical spectra obtained with the GRACES spectrograph on the Gemini-N telescope and derived stellar metallicities and stellar parameters from a sample of 170 Fe I and Fe II lines with atomic parameters obtained from \citet{Heiter2015}. We derived the abundances of the elements Al, Na, Ni, Cr, as well as the heavy-elements Ba (s-process element) and Eu (r-process element).

We observed 13 candidate member stars in the TriAnd overdensity. Seven stars were confirmed as members of this population through a kinematic analysis and computation of stellar orbits using GAIA DR2 (\citealt{gaia_dr2}) proper motions along with our measured radial velocities of the stars. 
We also evaluated the membership of 17 additional TriAnd candidates analyzed in the previous high-resolution studies of \citet{Bergemann2018} and \citet{Hayes2018}, finding that two of these stars are probably not TriAnd members given their proper motions and orbit eccentricities.

One of the results of this study is the confirmation that the TriAnd overdensity has low-metallicity stars: our TriAnd sample has metallicities ranging from [Fe/H] = $-$1.34 $\pm$ 0.12 to $-$0.78 $\pm$ 0.1 dex, in contrast with the mono-metallicity of the TriAnd sample analyzed in \citet{Bergemann2018} ($<$[Fe/H]$>$=-0.57 $\pm$ 0.08). Our sample also extends to lower metallicities than the APOGEE sample (\citealt{Hayes2018}; their lowest metallicity star has [Fe/H] $\approx$ -1.1 dex). 

We find that the TriAnd overdensity is a structure composed of stars having disk-like orbits and a unique chemical pattern that does not entirely resemble the full abundance pattern observed for the stars in the local Galactic disk, nor dwarf spheroidal galaxies. 
TriAnd stars in our sample exhibit differences in the abundance patterns of [Na/Fe], [Al/Fe], (marginally [Ni/Fe]), [Ba/Fe] and [Eu/Fe], when compared to the Milky Way trend.
In particular, the heavy-element abundance ratios of [Eu/Ba] indicate that TriAnd is distinct, having low [Eu/Ba] ratios for all stars in our study; similar low values of [Eu/Ba] ratios are also found in \citet{Bergemann2018}. The exception is the most metal-poor TriAnd star in our sample, for which the [Eu/Ba] ratio is higher.

It should be noted that those target stars found to be non-TriAnd members based on a kinematic analysis generally exhibit a distinct chemical behavior when compared to the confirmed TriAnd members; the non-TriAnd stars have a chemical pattern that most closely resembles the chemical pattern of the local disk or halo. In addition one of the non-members stars can be chemically tagged to dwarf galaxy population as its abundances completely agree with the dwarf spheroidal pattern.
The targets that ended up not being from TriAnd serve as surrogate comparison stars and play an important role in the validation of the abundance offsets. 
The fact that we can chemically tag the TriAnd versus the non-TriAnd stars boosts confidence that the abundance differences found for TriAnd are not due to systematic differences in the abundance analyses. 
For its metallicity, TriAnd has a chemical pattern that is distinct from any known Galactic population. However, the chemical pattern of field stars in the very distant galaxy has not yet been fully probed.

The complexity of the abundance pattern for stars in the TriAnd overdensity, combined with the low number of TriAnd stars observed using high-resolution spectroscopy to date, and the paucity of studies chemically characterizing the outer disk population of Milky Way, are the main obstacles in unequivocally establishing the origin of the TriAnd population. Despite these barriers, the results in this paper find differences in the chemical patterns of TriAnd and the local Galactic disk, having a pattern that is also different from that of dwarf spheroidals.

\acknowledgments

We thank Allyson Sheffield for providing the metallicity data used in \citet{Sheffield2014} and Adrian M. Price-Whelan for the helpful comments about the Astropy library. We thank Maria Bergemann for extensive discussions. KC thanks Kathryn Johnston, and Chris Hayes for discussions.
JVSS thanks FAPERJ proc. 202.756/2016. HDP, HJR-P and FA-F thank the Brazilian Agency CAPES for the financial support of this research. HDP thanks FAPESP proc. 2018/21250-9 
This work has made use of data from the European Space Agency (ESA) mission
{\it Gaia} (\url{https://www.cosmos.esa.int/gaia}), processed by the {\it Gaia}
Data Processing and Analysis Consortium (DPAC,
\url{https://www.cosmos.esa.int/web/gaia/dpac/consortium}). Funding for the DPAC
has been provided by national institutions, in particular the institutions
participating in the {\it Gaia} Multilateral Agreement.

\vspace{5mm}
Facilities: Gemini North: GRACES, ESO-Archive: VLT: UVES.


Softwares:{\tt Astropy} \citep{astropy, astropy2018}
          IRAF 
          Opera 
          {\tt matplotlib} \citep{matplolib},
          {\tt Numpy} \citep{numpy},
          {\tt Scipy} \citep{scipy}.

\appendix

\section{Abundance and atmospheric parameters uncertainties}

The uncertainty in the effective temperature was estimated from the uncertainty in the slope of the excitation potential versus abundance plot (which defined the excitation equilibrium) obtained by varying only the effective temperature until the slope increased by sigma. In the same way, the uncertainty in the slope of the equivalent width versus abundance plot defined the uncertainty in the microturbulence velocity. The log $g$ uncertainty was determined by varying the log $g$ until the abundance of Fe~II (which defined the ionization equilibrium) increased by sigma. We performed these uncertainty steps for the star 11 and used it as a reference for the other stars in our sample. Thus we define 75 K, 0.2 dex and 0.1 km/s as the uncertainties of T$_{\rm eff}$, log $g$ and $\xi$, respectively, for the stars of our sample.
   
To calculate the uncertainties in the abundances we first determined the uncertainties caused independently by each atmospheric parameter varying these parameters of their respective uncertainties. After this, we obtained the final uncertainties of chemical abundances adding quadratically the uncertainties in abundance relative to each atmospheric parameter. In Table \ref{tab:error} we show the uncertainties regarding the star 11. For the other stars in our sample we have similar uncertainties.

\begin{table*}[h]
\caption{Abundance uncertainties for star 11.} 
\centering
\label{tab:error}
\begin{tabular}{lcccc}\\\hline\hline
Element & $\Delta T_{eff}$ & $\Delta\log g$ & $\Delta\xi$ & $\left( \sum \sigma^2 \right)^{1/2}$ \\
$_{\rule{0pt}{8pt}}$ & $+$75~K & $+$0.2 & $+$0.1 km\,s$^{-1}$ &  \\
\hline     
Fe\,{\sc i}    & $-$0.01  & $+$0.05 & $-$0.05 & 0.07 \\ 
Fe\,{\sc ii}   & $-$0.13  & $+$0.12 & $-$0.02 & 0.18 \\
Na\,{\sc i}    & $+$0.07  & $+$0.01 & $-$0.03 & 0.08 \\
Al\,{\sc i}    & $+$0.05  & $+$0.01 & $-$0.02 & 0.05 \\
Cr\,{\sc i}    & $+$0.10  & $+$0.03 & $-$0.11 & 0.15 \\
Ni\,{\sc i}    & $-$0.02  & $+$0.06 & $-$0.03 & 0.07 \\
Ba\,{\sc ii}   & $+$0.04  & $+$0.10 & $-$0.10 & 0.15 \\
Eu\,{\sc ii}   & $-$0.01  & $+$0.10 & $-$0.02 & 0.10 \\
\hline
\end{tabular}

\par Notes. Each column gives the variation of the abundan-\\ce caused by the variation in $T_{\rm eff}$, $\log g$ and $\xi$. The last\\ column gives the compounded rms uncertainty of the se-\\cond to fourth columns.

\end{table*}

\section{Chemical abundances of the Sun and Arcturus} \label{sec:pubcharge}

   We also analyzed the Arcturus and Sun spectra to test our methodology and linelist. The atmospheric parameters obtained for Arcturus are very similar with the results derived by \citet{RamirezAllendePrieto2011}, with difference of 84 K for T$_{\rm eff}$, 0.04 for log $g$, 0.1 km/s for $\xi$ and 0.07 dex for [Fe/H]. For the Sun we determined T$_{\rm eff}$ = 5820 K, log $g$ = 4.5, and $\xi$ = 1.04 km/s. \citet{Smiljanic2014} tested all the methodologies used in the GAIA-ESO survey with star references, such as Arcturus. The difference between our atmospheric parameters for Arcturus is lower than the difference of the parameters obtained by several GAIA-ESO methodologies, giving reliability to our methodology. In Table \ref{tab:arcturus_sun} we show our results for the chemical abundance of all elements for Arcturus and the Sun. The chemical abundance for Na, Al, Ni, and r- and s-process elements for Arcturus in our study present similar values with the results of \citet{RamirezAllendePrieto2011}, with a difference smaller than 0.08 dex. \citet{RamirezAllendePrieto2011} did not perform NLTE corrections in Cr abundances that are of the order of 0.2 (dex) in Arcturus (\citealt{BergemannNordlander2014}), which justifies the difference in the abundance of this element between our results and those of \citet{RamirezAllendePrieto2011}. Our abundance results for the Sun also present values similar to those in the literature (Table \ref{tab:arcturus_sun}). We used our solar abundance to normalize the chemical abundances of all elements.

\begin{table*}[h]
\caption{Arcturus and solar abundances.}
\label{tab:arcturus_sun}
\tabcolsep 0.7truecm
\label{sun}
\begin{tabular}{lccc|cc}\hline\hline
\multicolumn{4}{c}{Sun} & \multicolumn{2}{c}{Arcturus}\\\hline
Element              & This  & Grevesse \&  & Asplund        & This  & Ramirez \&          \\
$_{\rule{0pt}{8pt}}$ & work  & Sauval (1998)& et al. (2009)  & work  & Allende Prieto (2011)\\
\hline     
Fe                   & 7.50  & 7.50         & 7.50           & 6.91  & 6.98                 \\ 
Na                   & 6.24  & 6.33         & 6.24           & 5.78  & 5.81                 \\
Al                   & 6.51  & 6.47         & 6.45           & 6.26  & 6.25                 \\
Cr                   & 5.86  & 5.67         & 5.64           & 5.24  & 4.99                 \\
Ni                   & 6.24  & 6.25         & 6.22           & 5.71  & 5.73                 \\
Ba                   & 2.25  & 2.13         & 2.18           & 1.53  & 1.50$^{a}$           \\
Eu                   & 0.59  & 0.51         & 0.52           & 0.33  & 0.25$^{b}$           \\
\hline
\end{tabular}
Notes. $^{a}$ Ba abundance from \citet{Smith2000}. $^{b}$ Eu abundance for 6645\AA{} line from \citet{Overbeek2016}.
\end{table*}

\section{stellar orbits}
\label{appendix:orbits}

In this appendix we present the stellar orbits integrated for the stars in our sample, and that were used to derive the results shown in Section \ref{sec:kinematics}. The orbits were integrated for 3.5 Gyr considering a Milky Way potential approximated as a combination of a power law with cut off bulge, a Miyamoto-Nagai disk, and a Navarro-Frenk-White halo. The integration was made using the orbital integrator {\tt galpy} \citep{Bovy2015}. 

The input parameters for the integration are the stellar position, Gaia's proper motion, our measured radial velocities through spectroscopy and our estimated distances through photometric parallax. The orbits were characterized in terms of eccentricity and orbital diskness (Equations \ref{eq:ecc} and \ref{eq:od}) in order to select the TriAnd members \ref{fig:ecc_disk}.

\begin{figure}
\centering
\includegraphics[width=0.95\columnwidth]{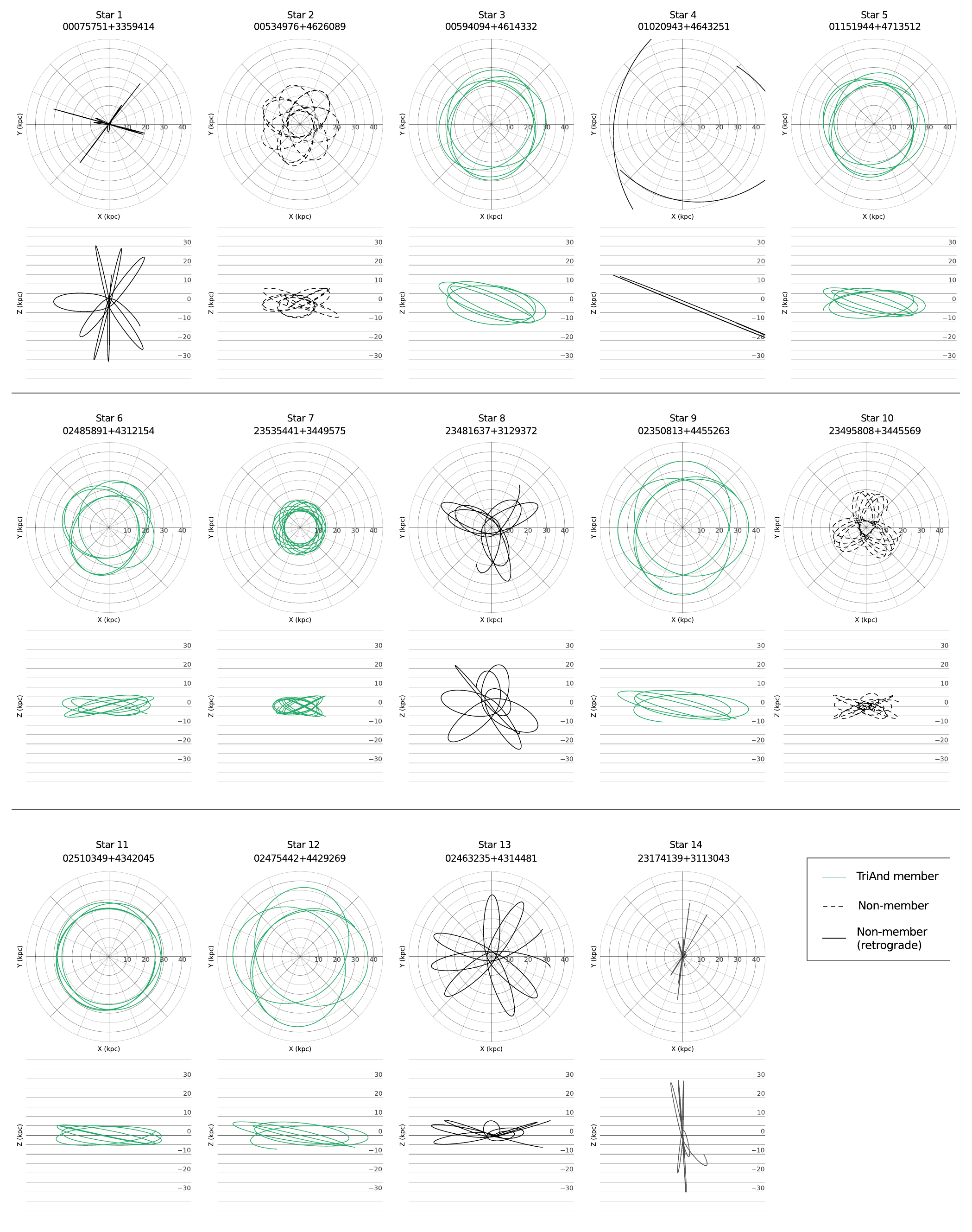}
\caption{Orbits integrated for all 13 stars in our sample shown in the XY and XZ plane. The orbits of stars classified as members of TriAnd are shown in green while the non-members are shown in black. The dashed line indicate that the orbit is retrograde.}
\label{fig:orbits}
\end{figure}

In Figure \ref{fig:orbits}, we show the orbits of all 13 stars in our sample, both in the XY disk plane, and in the XZ plane. The stars classified as members both by the proper motion and orbital parameters criteria are shown in green, and the black orbits corresponds to the stars classified as non-members by both methods (dashed lines further indicate that the orbit is retrograde).

\section{Equivalent width measurements} \label{sec:equivalentwidth}

\begin{table}
\caption{Observed Fe\,{\sc i} and Fe\,{\sc ii} lines.}
\scriptsize
\centering
\begin{tabular}{cccccccccccccccccc}\\\tableline\tableline
\label{tabelFeb}
        &            &         &          & \multicolumn{14}{c}{Equivalent Widths (m\AA)}
\\\tableline
        &            &         &          & \multicolumn{14}{c}{Star} \\
\cline{5 - 18}
Element &  $\lambda$\,(\AA) & $\chi$(eV) & log $gf$  & 1 & 2 & 3 & 4 & 5 & 6 & 7 & 8 & 9 & 10 & 11 & 12 & 13 & 14 \\
\tableline
Fe\,{\sc i} &   5125.12  &  4.22  &$-$0.140 &  --- & --- & ---&  69&  ---& ---  &  81 &--- & 75 & 146 & ---&  ---& ---&  ---\\ 
 &   5133.69  &  4.18  &   0.360 &  --- & --- & ---& ---&  ---& ---&  ---&  --- &  89 & ---& ---&  ---& ---&  ---\\
 &   5159.06  &  4.28  &$-$0.820 &  --- & --- & 102& ---&  ---& ---&  ---&  --- & --- & ---& ---&  ---& ---&   90\\
 &   5162.27  &  4.18  &   0.020 &  --- & 146 & ---& ---&  144& 142&  ---&  124 & --- & ---& 143&  ---& ---&  ---\\
 &   5242.49  &  3.63  &$-$0.967 &  --- & --- & ---& ---&  ---& ---&  ---&  105 & --- & ---& 120&  ---& ---&  115\\
 &   5250.21  &  0.12  &$-$4.933 &  --- & --- & ---& 131&  ---& ---&  117&  --- & --- & ---& ---&  132& ---&  ---\\
 &   5253.02  &  2.28  &$-$3.840 &  --- & 102 &  97&  40&  ---& ---&  ---&   63 & --- &  84&  81&  ---& ---&   84\\
 &   5288.52  &  3.69  &$-$1.493 &  --- & 117 & 113& ---&  ---&  97&  ---&  --- & --- & 115& ---&  ---&  95&   82\\
 &   5302.30  &  3.28  &$-$0.738 &  --- & --- & ---& ---&  ---& ---&  ---&  146 & --- & ---& ---&  ---& ---&  ---\\
 &   5315.07  &  4.37  &$-$1.650 &  --- &  51 & ---& ---&  ---& ---&  ---&   20 &  24 & ---&  52&   21&  34&   31\\
 &   5321.11  &  4.43  &$-$1.089 &  --- & --- &  77& ---&   74& ---&  ---&   36 & --- & ---&  61&   32& ---&  ---\\
 &   5322.04  &  2.28  &$-$2.802 &  --- & --- & ---& ---&  142& ---&  ---&  --- & --- & 144& ---&  ---& ---&  139\\
 &   5364.87  &  4.45  &   0.228 &  --- & --- & 133& ---&  133& ---&  ---&  116 & --- & ---& ---&  ---& 132&  116\\
 &   5367.47  &  4.42  &   0.444 &  --- & --- & ---& ---&  ---& ---&  ---&  --- & --- & ---& ---&  ---& ---&  121\\
 &   5369.96  &  4.37  &   0.536 &  --- & --- & ---& ---&  ---& ---&  ---&  130 & --- & ---& ---&  ---& ---&  134\\
 &   5373.71  &  4.47  &$-$0.710 &  --- & --- &  91&  37&   84& ---&  ---&   65 & --- &  89&  89&   54&  83&   79\\
 &   5389.48  &  4.42  &$-$0.410 &   47 & 119 & 119&  61&  109& 107&   71&  --- & --- & ---& ---&   63& ---&  ---\\
 &   5400.50  &  4.37  &$-$0.160 &   61 & --- & ---& ---&  ---& ---&  ---&  --- & --- & ---& ---&  ---& ---&  ---\\
 &   5410.91  &  4.47  &   0.398 &  --- & --- & 139& ---&  135& ---&  ---&  --- & --- & ---& ---&  ---& 138&  ---\\
 &   5417.03  &  4.42  &$-$1.580 &   13 &  55 &  47&  14&  ---&  43&   26&   24 &  25 &  47&  45&  ---&  36&   35\\
 &   5441.34  &  4.31  &$-$1.630 &  --- &  51 & ---&  13&  ---& ---&  ---&  --- & --- & ---&  54&   20& ---&  ---\\
 &   5445.04  &  4.39  &$-$0.020 &  --- & --- & ---& ---&  ---& ---&  ---&  114 & --- & 131& 117&  ---& 129&  113\\
 &   5487.74  &  4.32  &$-$0.317 &  --- & --- & ---&  59&  ---& ---&   83&  --- & --- & ---& ---&  ---& ---&  ---\\
 &   5522.45  &  4.21  &$-$1.450 &   20 & --- & ---&  28&   68& ---&  ---&   38 &  37 &  62&  61&   33&  61&   51\\
 &   5532.75  &  3.57  &$-$2.050 &  --- & --- & ---&  34&  ---& ---&   59&  --- &  49 &  96& ---&  ---& ---&   85\\
 &   5554.89  &  4.55  &$-$0.270 &   46 & 120 & 115&  47&  106& 111&   71&   78 &  61 & ---& ---&   59& 103&   97\\
 &   5560.21  &  4.43  &$-$1.090 &   19 &  70 & ---&  26&   81& ---&  ---&  --- &  33 & ---&  61&  ---&  67&   55\\
 &   5567.39  &  2.61  &$-$2.568 &  --- & --- & ---&  71&  139& 146&   76&  --- &  70 & ---& ---&   82& ---&  ---\\
 &   5576.09  &  3.43  &$-$0.900 &  --- & --- & ---& ---&  ---& ---&  ---&  --- & --- & ---& 140&  ---& ---&  141\\
 &   5584.77  &  3.57  &$-$2.220 &  --- &  91 & ---&  27&  ---&  86&  ---&   46 & --- & ---&  66&  ---&  82&   68\\
 &   5624.02  &  4.39  &$-$1.380 &  --- &  73 & ---& ---&  ---& ---&   45&  --- & --- &  68& ---&  ---& ---&   52\\
 &   5633.95  &  4.99  &$-$0.230 &   23 &  79 &  86&  37&   69&  77&   47&  --- &  42 &  74&  70&   42&  75&   67\\
 &   5635.82  &  4.26  &$-$1.790 &   14 &  57 & ---& ---&  ---&  52&  ---&   25 &  27 &  54& ---&  ---& ---&  ---\\
 &   5638.26  &  4.22  &$-$0.720 &   43 & 111 & 108& ---&  111& 103&   65&   78 &  57 & 107& 102&  ---& 114&   92\\
 &   5658.82  &  3.40  &$-$0.766 &   93 & --- & ---& ---&  ---& ---&  ---&  --- &  99 & ---& ---&  ---& ---&  ---\\
 &   5686.53  &  4.55  &$-$0.455 &   39 & --- & 102&  42&  104&  87&  ---&   74 &  53 & ---&  87&   53&  88&   77\\
 &   5691.50  &  4.30  &$-$1.450 &   22 &  69 &  78&  16&   65&  56&   42&  --- &  33 &  69&  67&   35& ---&   49\\
 &   5705.46  &  4.30  &$-$1.355 &   18 &  71 &  62&  25&  ---&  69&  ---&   44 & --- &  70& ---&   36& ---&   44\\
 &   5717.83  &  4.28  &$-$0.990 &   30 & --- & ---& ---&  ---& ---&  ---&  --- & --- & ---&  83&  ---&  78&   64\\
 &   5731.76  &  4.26  &$-$1.200 &  --- & --- &  85& ---&  ---& ---&  ---&  --- &  39 &  80&  77&  ---& ---&   61\\
 &   5762.99  &  4.21  &$-$0.360 &  --- & --- & 132&  66&  130& 137&   84&  112 &  72 & ---& ---&   85& 134&  ---\\
 &   5806.73  &  4.61  &$-$0.950 &   24 &  74 &  70& ---&   66&  60&   48&   45 &  38 &  74&  69&   42&  69&  ---\\
 &   5814.81  &  4.28  &$-$1.870 &  --- &  53 &  47&  11&   49&  54&   26&  --- & --- &  50&  40&   25& ---&  ---\\
 &   5883.82  &  3.96  &$-$1.260 &  --- & 100 &  96& ---&  105& ---&  ---&   77 & --- & ---&  91&  ---& ---&   87\\
 &   5916.25  &  2.45  &$-$2.994 &  59  & 133 & 138&  73&  129& 135&   79&  114 &  72 & 136& 125&   79& 123&  113\\
 &   5934.65  &  3.93  &$-$1.070 &  54  & --- & 116& ---&  126& 119&   75&   96 &  69 & ---& 111&  ---& 110&  109\\
 &   6016.60  &  3.55  &$-$1.720 &  49  & --- & ---&  65&  ---& ---&  ---&  --- & --- & ---& ---&  ---& ---&  ---\\
 &   6020.17  &  4.61  &$-$0.270 &  42  & --- & ---& ---&  ---& ---&  ---&  --- & --- & ---& ---&  ---& ---&  ---\\
 &   6024.06  &  4.55  &$-$0.120 &  62  & 124 & ---&  57&  118& ---&  ---&  --- & --- & ---& 114&  ---& 117&  103\\
 &   6027.05  &  4.08  &$-$1.089 &  --- & 110 & ---&  48&   93& ---&  ---&   83 & --- & ---&  90&  ---&  98&   79\\
 &   6056.01  &  4.73  &$-$0.320 &  --- & --- & ---&  38&  ---& ---&  ---&   63 & --- & ---& ---&   53&  89&  ---\\
 &   6065.48  &  2.61  &$-$1.529 &  112 & --- & ---& 128&  ---& ---&  127&  --- & 118 & ---& ---&  127& ---&  ---\\
 &   6079.01  &  4.65  &$-$1.020 &   21 & --- &  65&  21&   65&  53&   40&   32 &  37 &  53&  56&   39&  60&   48\\
 &   6082.71  &  2.22  &$-$3.576 &   57 & --- & 114&  68&  ---& 133&   74&   97 &  70 & ---& 113&   81& ---&  111\\
 &   6093.64  &  4.61  &$-$1.400 &  --- &  60 &  48& ---&   36& ---&   31&   25 &  24 &  43&  41&  ---& ---&  ---\\
 &   6096.66  &  3.98  &$-$1.830 &   18 &  68 &  75&  31&   75&  68&   45&  --- &  41 &  64&  63&  ---&  66&   50\\
 &   6120.25  &  0.92  &$-$5.970 &   33 & 101 &  92&  47&   98& 105&   55&   63 &  52 & 100&  88&   66&  88&   94\\
 &   6136.61  &  2.45  &$-$1.402 &  126 & --- & ---& 149&  ---& ---&  ---&  --- & 146 & ---& ---&  ---& ---&  ---\\
\tableline
\end{tabular}
\end{table} 
 
\begin{table}
\noindent
\scriptsize
\centering
\begin{tabular}{cccccccccccccccccc}
\multicolumn{18}{c}{Table 6, continued.}\\
\tableline\tableline
        &            &         &          & \multicolumn{14}{c}{Equivalent Widths (m\AA)}
\\\tableline
        &            &         &          & \multicolumn{14}{c}{Star} \\
\cline{5 - 18}
Element &  $\lambda$\,(\AA) & $\chi$(eV) & log $gf$  & 1 & 2 & 3 & 4 & 5 & 6 & 7 & 8 & 9 & 10 & 11 & 12 & 13 & 14 \\
\tableline
Fe\,{\sc i}  &   6151.62  &  2.18  &$-$3.295 &   67 & 141 & 141&  82&  146& 149&   79&  129 &  80 & 148& 135&   85& 137&  136\\
 &   6157.73  &  4.08  &$-$1.160 &   44 & --- & ---&  53&  111& ---&  ---&  --- & --- & ---& ---&  ---& ---&  ---\\
&   6165.36  &  4.14  &$-$1.473 &   28 &  84 &  81&  28&   76&  82&   52&   50 &  36 & ---&  70&   37&  67&   60\\
 &   6173.33  &  2.22  &$-$2.880 &  78  & --- & ---& ---&  ---& ---&   94&  --- &  86 & ---& 140&  ---& ---&  ---\\
 &   6187.99  &  3.94  &$-$1.620 &  --- & 100 &  86&  35&   77& ---&  ---&  --- & --- &  75&  85&   43&  72&   74\\
 &   6191.56  &  2.43  &$-$1.416 &  --- & --- & ---& 135&  ---& ---&  146&  --- & --- & ---& ---&  ---& ---&  ---\\
 &   6200.31  &  2.61  &$-$2.433 &  --- & --- & ---& ---&  144& ---&  ---&  133 & --- & 149& 131&  ---& 147&  131\\
 &   6213.43  &  2.22  &$-$2.481 &  --- & --- & ---& 112&  ---& ---&  108&  --- & 103 & ---& ---&  ---& ---&  ---\\
 &   6252.56  &  2.40  &$-$1.699 &  121 & --- & ---& ---&  ---& ---&  147&  --- & 132 & ---& ---&  140& ---&  ---\\
 &   6254.26  &  2.28  &$-$2.439 &  --- & --- & ---& ---&  ---& ---&  ---&  --- & --- & ---& ---&  123& ---&  ---\\
 &   6311.50  &  2.83  &$-$3.141 &  --- & --- & ---&  49&   92&  86&   60&   67 &  48 & ---&  76&   51&  93&   81\\
 &   6322.69  &  2.59  &$-$2.430 &  --- & --- & ---&  99&  ---& 146&  ---&  140 & --- & ---& 145&  105& 144&  134\\
 &   6380.74  &  4.19  &$-$1.375 &   25 &  91 &  79&  30&   85& ---&   48&   58 &  40 &  89&  82&   43&  70&   75\\
 &   6393.60  &  2.43  &$-$1.452 &  122 & --- & ---& 146&  ---& ---&  ---&  --- & 139 & ---& ---&  150& ---&  ---\\
 &   6411.65  &  3.65  &$-$0.596 &   89 & --- & ---& 105&  ---& ---&  107&  --- &  98 & ---& ---&  112& ---&  140\\
&   6419.95  &  4.73  &$-$0.200 &   45 & 101 & 103&  59&  103& ---&   67&  --- &  57 &  99&  90&   68& ---&   80\\
 &   6436.41  &  4.19  &$-$2.580 &  --- & --- & ---& ---&  ---& ---&   15&  --- & --- &  17& ---&  ---& ---&  ---\\
 &   6518.37  &  2.83  &$-$2.438 &  67  & --- & 130&  82&  130& 130&   83&  112 &  75 & 128& 117&   90& 130&  117\\
&   6551.68  &  0.99  &$-$5.790 &  36  & --- & 108& ---&  ---& ---&   52&   94 &  55 & ---& ---&   61& ---&  107\\
 &   6574.23  &  0.99  &$-$5.004 &  --- & --- & ---& ---&  ---& ---&  ---&  147 & --- & ---& ---&  ---& ---&  ---\\
&   6592.91  &  2.73  &$-$1.473 &  --- & --- & ---& 131&  ---& ---&  141&  --- & --- & ---& ---&  137& ---&  ---\\
 &   6597.56  &  4.80  &$-$0.970 &  16  &  57 &  54&  23&   44&  63&   32&   30 & --- &  55&  57&  ---&  46&   40\\
 &   6608.02  &  2.28  &$-$3.930 &  35  & --- & 106&  51&   86& 115&  ---&   64 & --- & 102&  95&   64&  98&   81\\
 &   6609.11  &  2.56  &$-$2.691 &  --- & --- & ---& ---&  ---& ---&  ---&  124 & --- & 145& ---&  ---& ---&  136\\
 &   6653.85  &  4.15  &$-$2.215 &  --- & --- & ---&  10&   29&  31&  ---&   14 & --- &  32& ---&   13&  24&  ---\\
 &   6699.14  &  4.59  &$-$2.110 &  --- &  20 &  19& ---&  ---&  17&   13&  --- & --- &  12&  13&    9& ---&  ---\\
 &   6703.57  &  2.76  &$-$3.060 &  49  & 118 &  99&  63&  101& 108&   68&   82 & --- & 104&  97&   72&  94&   91\\
 &   6704.48  &  4.22  &$-$2.380 & ---  &  26 & ---& ---&   17&  31&  ---&  --- & --- & ---& ---&   11&  12&   25\\
 &   6710.32  &  1.49  &$-$4.764 &  50  & --- & ---&  67&  122& 119&   74&   96 &  70 & 114& 111&   80& 109&  105\\
 &   6713.74  &  4.80  &$-$1.500 &  --- & --- & ---& ---&  ---& ---&   23&  --- &  16 & ---& ---&  ---&  23&  ---\\
 &   6739.52  &  1.56  &$-$4.794 &  37  & 113 & ---& ---&  ---& 115&  ---&  --- & --- & ---& ---&  ---& ---&   97\\
 &   6745.96  &  4.08  &$-$2.500 & ---  & --- &  33& ---&   19& ---&  ---&  --- & --- & ---& ---&   11& ---&  ---\\
 &   6750.15  &  2.42  &$-$2.618 &  81  & --- & ---&  96&  ---& ---&   92&  147 & --- & ---& 149&  ---& ---&  ---\\
 &   6783.70  &  2.59  &$-$3.980 &   20 & --- &  77&  22&  ---&  62&   37&   46 & --- &  57& ---&  ---& ---&   49\\
 &   6793.26  &  4.08  &$-$2.326 &  --- & --- &  36&  13&   30& ---&   28&  --- &  14 & ---&  32&  ---&  28&  ---\\
 &   6810.26  &  4.61  &$-$0.986 &   25 &  85 &  69&  30&   70& ---&   47&   40 & --- & ---&  62&   45&  62&  ---\\
 &   6820.37  &  4.64  &$-$1.220 &  --- & --- &  58& ---&  ---& ---&  ---&   34 & --- & ---& ---&  ---& ---&  ---\\
 &   6841.34  &  4.61  &$-$0.490 &   38 & --- &  98&  49&  ---& ---&  ---&  --- &  55 &  96& ---&   56& ---&  ---\\
 &   6851.64  &  1.61  &$-$5.320 &  --- &  77 &  70& ---&   64&  71&   39&   42 &  42 &  75&  65&  ---&  62&  ---\\
 &   6858.15  &  4.61  &$-$0.903 &   31 &  74 &  68&  34&   61&  72&   50&   41 & --- &  66&  64&  ---&  64&  ---\\
 &   7130.92  &  4.22  &$-$0.690 &   59 & --- & 130&  75&  127& ---&  ---&   93 & --- & ---& 122&  ---& 117&  ---\\
 &   7132.99  &  4.08  &$-$1.628 &   22 &  86 &  70&  39&   71& ---&   48&   45 & --- &  82&  74&  ---&  67&  ---\\
Fe\,{\sc ii}  &  5234.62  &  3.22  &$-$2.180 & --- &  76& 103&  ---&  91&  81 & ---&  --- & --- &  83& ---&  ---&  88&   73\\
  &  5276.00  &  3.20  &$-$1.940 &  65 & --- & ---& ---&  ---& ---&  ---&  --- & --- & ---& ---&  ---& ---&  ---\\
  &  5284.10  &  2.89  &$-$3.195 &  39 &  53 & ---& ---&  ---&  58&  ---&   52 & --- &  47&  53&  ---&  60&   56\\
  &  5325.55  &  3.22  &$-$3.160 & --- & --- & ---& ---&   41& ---&  ---&  --- & --- & ---&  42&  ---& ---&   32\\
  &  5414.07  &  3.22  &$-$3.580 & --- & --- & ---& ---&  ---& ---&  ---&  --- & --- & ---&  21&  ---& ---&  ---\\
  &  5425.25  &  3.20  &$-$3.220 &  21 &  31 & ---& ---&  ---&  38&  ---&  --- &  26 &  26&  39&  ---&  52&   30\\
  &  5534.84  &  3.25  &$-$2.865 &  31 & --- & ---&  32&  ---& ---&  ---&   50 & --- & ---& ---&  ---& ---&  ---\\
  &  5991.37  &  3.15  &$-$3.647 & --- & --- &  38&  20&  ---& ---&   23&   20 &  22 & ---& ---&  ---& ---&   25\\
  &  6084.10  &  3.20  &$-$3.881 &   8 & --- &  28&  12&   25& ---&   15&  --- &  15 &   9&  17&  ---&  23&  ---\\
  &  6149.25  &  3.89  &$-$2.841 &  13 & --- & ---&  18&  ---& ---&   18&   15 & --- & ---&  26&   16&  23&  ---\\
  &  6247.56  &  3.89  &$-$2.435 & --- &  27 & ---&  29&   46&  31&   30&  --- &  24 & ---&  35&   30&  49&  ---\\
  &  6416.92  &  3.89  &$-$2.877 & --- & --- &  39& ---&  ---& ---&  ---&  --- & --- & ---& ---&  ---& ---&  ---\\
  &  6432.68  &  2.89  &$-$3.570 &  28 &  36 &  55&  30&   40&  45&   33&   33 &  34 & ---&  39&   29&  41&   30\\
\tableline
\end{tabular}
\end{table}
 
\begin{table*}
\caption{Na, Al, Cr and Ni lines analyzed.}
\scriptsize
\centering
\begin{tabular}{cccccccccccccccccc}
\label{tabellinesx}
\\\tableline\tableline
    & & & &\multicolumn{14}{c}{Equivalent Widths (m\AA)} \\\tableline
\multicolumn{4}{c}{} & \multicolumn{14}{c}{Star}\\
\cline{5-18}
Element & $\lambda$ & $\chi$(eV) & $\log gf$ & 1 & 2 & 3 & 4 & 5 & 6 & 7 & 8 & 9 & 10 & 11 & 12 & 13 & 14 \\
\tableline
Na\,{\sc i} & 5682.63 & 2.10 & $-$0.700 &  36 & --- & --- &  48 & --- & --- & --- &  85 &  86 & 126 & --- &  93 & 130 & --- \\
Na\,{\sc i} & 5688.20 & 2.10 & $-$0.400 &  52 & --- & --- &  59 & --- & --- & --- &  92 &  90 & 133 & --- &  94 & 140 & --- \\
Na\,{\sc i} & 6154.22 & 2.10 & $-$1.510 &   9 & 103 &  98 &  16 &  84 & 104 &  42 &  23 &  53 &  56 &  86 &  52 &  57 &  36 \\
Na\,{\sc i} & 6160.75 & 2.10 & $-$1.210 &  14 & 121 & 110 &  25 & 120 & 112 &  52 &  38 &  57 &  77 & 104 &  64 &  72 &  56 \\
Al\,{\sc i} & 6696.02 & 3.14 & $-$1.570 &  14 & 106 &  88 &  19 &  86 & 107 &  59 & --- &  50 &  60 &  88 &  57 &  63 &  54 \\
Al\,{\sc i} & 6698.67 & 3.14 & $-$1.870 &  11 &  67 &  58 & --- &  53 &  75 &  27 & --- &  37 &  43 &  58 &  36 &  40 &  33 \\
Al\,{\sc i} & 7835.31 & 4.02 & $-$0.650 & --- &  82 &  66 & --- &  58 &  74 &  39 & --- &  46 &  37 &  63 &  45 &  41 & --- \\
Al\,{\sc i} & 8772.86 & 4.02 & $-$0.170 & --- & --- & --- &  25 & 103 & 117 &  66 & --- &  70 &  62 &  86 &  66 & --- & --- \\
Cr\,{\sc i} & 6330.10 & 0.94 & $-$2.920 &  52 & --- & --- &  66 & --- & --- &  88 & 119 &  88 & --- & 136 &  96 & 136 & 143 \\
Ni\,{\sc i} & 4913.97 & 3.74 & $-$0.500 & --- &  84 & --- & --- & 103 &  84 & --- & --- & --- & --- &  78 & --- & --- &  57 \\
Ni\,{\sc i} & 5084.10 & 3.68 & $-$0.084 & --- & 119 & 107 & --- & 115 & --- & --- & --- & --- & --- & --- & --- & 100 &  91 \\
Ni\,{\sc i} & 5094.41 & 3.83 & $-$0.998 & --- & --- & --- & --- & --- & --- & --- &  30 & --- & --- & --- & --- & --- & ---\\
Ni\,{\sc i} & 5115.39 & 3.83 & $-$0.110 & --- & 101 & --- & --- & --- & --- & --- &  82 & --- & --- & --- & --- & --- &  67 \\
Ni\,{\sc i} & 5157.98 & 3.61 & $-$1.510 &  11 &  47 &  34 &  14 & --- & --- & --- &  29 & --- &  34 &  47 & --- &  24 &  27 \\
Ni\,{\sc i} & 5589.36 & 3.90 & $-$0.938 &  11 &  59 &  68 &  19 & --- & --- & --- & --- & --- & --- &  49 & --- &  49 &  28 \\
Ni\,{\sc i} & 5593.73 & 3.90 & $-$0.682 & --- &  72 & --- & --- &  85 &  78 & --- &  35 & --- & --- &  66 & --- & --- &  40 \\
Ni\,{\sc i} & 5748.35 & 1.68 & $-$3.240 & --- & 126 & --- &  54 & 123 & --- &  60 & --- &  56 & --- & 113 & --- & 108 & --- \\
Ni\,{\sc i} & 5760.83 & 4.11 & $-$0.885 &   8 &  59 &  50 &  22 &  56 &  54 &  28 &  18 &  21 &  44 &  51 &  25 &  37 & --- \\
Ni\,{\sc i} & 5805.22 & 4.17 & $-$0.579 &  12 &  62 &  55 &  21 &  65 &  49 &  39 &  34 &  32 &  42 &  48 &  34 &  35 & --- \\
Ni\,{\sc i} & 5996.73 & 4.24 & $-$1.037 &   6 &  29 & --- &   6 &  27 & --- &  19 & --- & --- & --- &  35 &  19 & --- & --- \\
Ni\,{\sc i} & 6053.69 & 4.24 & $-$1.156 & --- & --- & --- & --- & --- & --- & --- &  14 & --- & --- &  31 & --- & --- & ---\\
Ni\,{\sc i} & 6086.28 & 4.27 & $-$0.410 &  16 &  55 &  62 &  23 &  67 &  53 &  36 &  25 &  35 &  44 & --- &  31 &  48 &  27 \\
Ni\,{\sc i} & 6108.12 & 1.68 & $-$2.600 &  82 & --- & --- &  98 & --- & --- & --- & 142 & --- & --- & --- & --- & --- & --- \\
Ni\,{\sc i} & 6111.07 & 4.09 & $-$0.865 &  10 &  52 &  54 &  19 &  56 & --- &  38 & --- &  32 &  44 &  53 &  34 &  35 & --- \\
Ni\,{\sc i} & 6128.97 & 1.68 & $-$3.430 &  49 & --- & --- &  61 & --- & 113 & --- &  93 & --- & 108 & --- &  77 &  99 & --- \\
Ni\,{\sc i} & 6176.81 & 4.09 & $-$0.260 &  32 &  81 &  78 &  37 &  80 &  83 &  53 &  57 &  49 &  65 &  78 & --- &  91 &  53 \\
Ni\,{\sc i} & 6177.24 & 1.83 & $-$3.460 &  32 &  80 &  79 &  36 & --- & 104 &  53 &  60 &  45 &  92 &  87 & --- &  63 &  65 \\
Ni\,{\sc i} & 6186.71 & 4.11 & $-$0.880 & --- & --- & --- & --- &  58 & --- &  25 &  26 &  27 & --- &  55 & --- & --- & --- \\
Ni\,{\sc i} & 6204.60 & 4.09 & $-$1.080 & --- & --- &  43 & --- & --- & --- & --- & --- &  21 & --- &  42 & --- & --- & --- \\
Ni\,{\sc i} & 6223.98 & 4.11 & $-$0.910 &  14 & --- &  54 &  18 &  44 &  40 &  27 &  21 &  23 &  40 &  39 &  37 & --- &  34 \\
Ni\,{\sc i} & 6230.09 & 4.11 & $-$1.260 & --- & --- &  31 &   8 & --- & --- &  17 &  16 &  20 & --- &  39 &  25 & --- & --- \\
Ni\,{\sc i} & 6322.17 & 4.15 & $-$1.115 & --- &  30 &  35 &   9 & --- & --- &  21 &  14 &  24 &  25 &  31 &  23 &  24 & --- \\
Ni\,{\sc i} & 6378.25 & 4.15 & $-$0.820 &   8 & --- &  55 &  19 & --- & --- &  29 &  19 & --- & --- &  59 &  30 &  33 & --- \\
Ni\,{\sc i} & 6384.66 & 4.15 & $-$1.130 & --- & --- & --- & --- & --- & --- &  21 & --- & --- & --- & --- & --- & --- & --- \\
Ni\,{\sc i} & 6482.80 & 1.94 & $-$2.630 &  57 & 136 & 132 &  84 & 128 & 139 & --- & 106 & --- & 122 & 125 & --- & 123 & 105 \\
Ni\,{\sc i} & 6532.87 & 1.94 & $-$3.350 &  32 &  99 &  96 &  44 &  96 & --- &  51 &  63 &  45 &  86 &  79 & --- &  84 &  73 \\
Ni\,{\sc i} & 6586.31 & 1.95 & $-$2.780 &  55 & 126 & 135 &  75 & 123 & 125 & --- &  93 & --- & 113 & 120 &  90 & 112 & 103 \\
Ni\,{\sc i} & 6598.60 & 4.24 & $-$0.821 & --- &  50 &  36 &  14 &  38 & --- &  26 & --- &  22 & --- &  38 &  34 &  37 & --- \\
Ni\,{\sc i} & 6635.12 & 4.42 & $-$0.765 & --- &  44 &  28 & --- &  31 & --- &  22 &  13 &  21 &  25 &  35 & --- &  28 &  21 \\
Ni\,{\sc i} & 6772.31 & 3.66 & $-$0.797 &  28 &  87 &  88 &  46 & --- & --- &  61 &  53 &  58 &  72 & --- &  63 & --- &  73 \\
Ni\,{\sc i} & 6842.04 & 3.66 & $-$1.374 &  20 & --- &  61 &  31 &  68 & --- & --- & --- &  38 & --- & --- & --- &  56 & --- \\
\tableline
\end{tabular}
\end{table*}



\end{document}